\documentclass[useAMS,usenatbib]{mnras}
\usepackage[T1]{fontenc} 
\usepackage{aecompl}
\usepackage{amsmath}
\usepackage{amssymb}
\usepackage{graphicx}
\usepackage{booktabs}
\usepackage{deluxetable}

\def\refjnl#1{{\rm#1}}
\def\araa{\refjnl{ARA\&A}}             
\def\apj{\refjnl{ApJ}}                 
\def\apjl{\refjnl{ApJ}}                
\def\apjs{\refjnl{ApJS}}               
\def\aap{\refjnl{A\&A}}                
\def\aaps{\refjnl{A\&AS}}              
\def\mnras{\refjnl{MNRAS}}             
\def\pasp{\refjnl{PASP}}               
\def\rmxaa{\refjnl{Rev. Mexicana Astron. Astrofis.}}

\newcommand{\HII}{H$\, \scriptstyle\rm II\ $} 
\newcommand{\CII}{C$\, \scriptstyle\rm II$} 
\newcommand{\OIII}{O$\, \scriptstyle\rm III$}

\newcommand{\simgt}{\lower.5ex\hbox{\gtsima}} 
\newcommand{\simlt}{\lower.5ex\hbox{\ltsima}} 
\newcommand{\simpr}{\lower.5ex\hbox{\prosima}} \def\la{\lsim} \def\ga{\gsim} 
\newcommand{\gtsima}{$\; \buildrel > \over \sim \;$} 
\newcommand{\ltsima}{$\; \buildrel < \over \sim \;$} 
 
\newcommand{\cloudy}{{\small CLOUDY}}

\title[GMC photoevaporation and FIR line emission]{Molecular clouds photoevaporation and FIR line emission}
\author[Vallini et al.]{L. Vallini$^{1,2,3}$\thanks{E-mail: livia.vallini@su.se}, A. Ferrara$^{4, 5}$, A. Pallottini$^{4,6,7}$, S. Gallerani$^{4}$\\
$^1$Nordita, KTH Royal Institute of Technology and Stockholm University, Roslagstullsbacken 23, SE-10691 Stockholm, Sweden\\
$^2$Dipartimento di Fisica e Astronomia, viale Berti Pichat 6, I-40127 Bologna, Italy \\
$^3$INAF, Osservatorio Astronomico di Bologna, via Ranzani 1, I-40127 Bologna, Italy \\
$^4$ Scuola Normale Superiore, Piazza dei Cavalieri 7, I-56126, Pisa, Italy\\
$^5$Kavli IPMU (WPI), Todai Institutes for Advanced Study, the University of Tokyo, Japan\\
$^{6}$Kavli Institute for Cosmology, University of Cambridge, Madingley Road, Cambridge CB3 0HA, UK\\
$^{7}$Cavendish Laboratory, University of Cambridge, 19 J. J. Thomson Ave., Cambridge CB3 0HE, UK}
\date{}
\pubyear{2011}

\begin{document}
\label{firstpage}
\pagerange{\pageref{firstpage}--\pageref{lastpage}} 
\maketitle
\begin{abstract}
With the aim of improving predictions on far infrared (FIR) line emission from Giant Molecular Clouds (GMC), we study the effects  of photoevaporation (PE) produced by external far-ultraviolet (FUV) and ionizing (extreme-ultraviolet, EUV) radiation on GMC structure. 
We consider three different GMCs with mass in the range $M_{\rm GMC} = 10^{3-6}\,\rm{M_{\odot}}$. Our model includes: (i) an observationally-based inhomogeneous GMC density field,  and (ii) its time evolution during the PE process. 
In the fiducial case ($M_{\rm GMC}\approx10^5 M_{\sun}$), the photoevaporation time ($t_{pe}$) increases from 1 Myr to 30 Myr for gas metallicity $Z=0.05-1\,\rm Z_{\odot}$, respectively.
Next, we compute the time-dependent luminosity of key FIR lines tracing the neutral and ionized gas layers of the GMCs, ([\CII] at $158\,\rm{\mu m}$, [\OIII] at $88\,\rm \mu m$) as a function of $G_0$, and $Z$ until complete photoevaporation at $t_{pe}$. We find that the specific [\CII] luminosity is almost independent on the GMC model within the survival time of the cloud. 
Stronger FUV fluxes produce higher [\CII] and [\OIII] luminosities, however lasting for progressively shorter times. At $Z=Z_{\odot}$ the [\CII] emission is maximized ($L_{\rm CII}\approx 10^4\,\rm{L_{\odot}}$ for the fiducial model) for $t<1\,\rm{Myr}$ and $\log G_0\geq 3$. Noticeably, and consistently with the recent detection by Inoue et al. (2016) of a galaxy at redshift $z\approx 7.2$, for $Z\leq 0.2\,\rm{Z_{\odot}}$ the [\OIII] line might outshine [\CII] emission by up to $\approx 1000$ times. We conclude that the [\OIII] line is a key diagnostic of low metallicity ISM, especially in galaxies with very young stellar populations.
\end{abstract}
\section{Introduction}
Giant molecular clouds (GMCs) are the reservoirs of molecular gas fuelling the star formation (SF) in galaxies. The complex network of physical processes linking the SF with the global evolution of a GMC it is often referred to as \textit{feedback} 
\citep[see][and references therein.]{mckee2007} 
Feedback determines the rate at which GMCs eventually return their gas to the diffuse phase of the interstellar medium (ISM), hence setting the efficiency of the subsequent episodes of star formation which thus are self-regulated. 
The feedback processes acting on GMCs scales are ultimately due to radiative and/or mechanical energy injection, both within and from outside the clouds.

External momentum, can be provided by supernova (SN) explosions and superbubbles \citep[e.g.][]{wada2001, elmegreen2004}, and by spiral shocks \citep[][]{bonnell2006}. 
Internally, the mechanical energy is provided by jet/outflows/winds from protostars and newly formed stars \citep[e.g.][]{norman1980, dale2013, nakamura2014}, SN explosions \citep[e.g.][]{walch2015, kim2016, kortgen2016}, and expanding \HII regions powered by newborn star clusters within the cloud \citep[e.g.][]{bertoldi1989, williams1997, krumholz2006, vazquez-semadeni2010, dale2014}.
More precisely the ionizing photons produced by young massive stars forming into (or close by) GMCs produce \HII regions that, by expanding into the ambient gas, energize the molecular clouds, thus contributing to the large-scale turbulent power \citep[][and references therein]{elmegreen2004}. Supersonic turbulence inhibits the collapse of GMCs and regulate the SF efficiency \citep[][]{maclow2004, vazquez-semadeni2005, krumholz2005, padoan2014}. In spite of this, the net effect of star formation is negative,  and ultimately photoevaporates and unbinds GMCs in a few dynamical times \citep{bertoldi1989, bertoldi1990, gorti2002, krumholz2006}.

Beside extreme UV (EUV), ionizing \textbf{($h \nu<13.6\,\rm{eV}$)} photons, also far-ultraviolet (FUV) photons ($6\,\rm{eV}<h \nu<13.6\,\rm{eV}$) strongly affect the chemistry, thermal balance, structure, and dynamics of GMCs. FUV radiation dissociates molecular gas beyond the \HII region, creating photodissociation regions \citep[PDRs,][]{hollenbach1999} from which most infrared (IR) emission of galaxies originates. Dust grains and polycyclic aromatic hydrocarbons, absorb radiation from the stars and reradiate this energy in the IR; at the same time photoelectrons heat the gas \citep{wolfire2003}. Radiative cooling is enabled by many far infrared (FIR) lines. The importance of FIR line emission in constraining the ISM properties (e.g. gas temperature, density, and metallicity) has driven the advances in the IR and sub-millimeter astronomy. Nowadays, with the Atacama Large Millimeter/Submillmeter Array (ALMA) we can aim at constraining the properties of PDRs and those of the associated molecular clouds in the first galaxies. One of the ALMA primary goals is the detection of the (redshifted) [\CII] from the high-$z$ Universe \citep[e.g.][]{carilli2013} even though the physical interpretation of the line measurement from high redshift is often very challenging \citep[][]{maiolino2015, gallerani2016, knudsen2016}. Many recent efforts \citep{nagamine2006, vallini2013, vallini2015, olsen2015, pallottini2015, pallottini2017, gallerani2016} have been devoted to relate the physical properties of neutral and molecular gas in high-$z$ galaxies to the FIR line luminosity. However,radiative feedback effects have not yet been included in such relation. Here we aim at making this step. 

Our plan is to model the time evolution of FIR line emission from a single GMC illuminated by nearby massive stars. Our investigation builds upon previous studies \citep{bertoldi1989, bertoldi1990, gorti2002} on GMC photoevaporation (PE) induced by externally-produced EUV/FUV photons impinging on the cloud. We adopt their analytic formalism to model the PE process and compute the time-evolution of the GMC density field. Then, by coupling the model with the photoionization/photodissociation code \cloudy, we self-consistently calculate the evolving luminosity of several FIR lines. Among these, the [\CII] 158$\mu$m fine structure line is often the most luminous and it is considered the work-horse for high-$z$ galaxy exploration. We focus our attention also on [\OIII] at $88\, \rm{\mu m}$, originating from the outer, ionized shell. The final goal is to assess whether (and how) PE, by affecting the density field of the GMCs, modulates FIR line emission under a range of different metallicity and irradiation conditions.  

The paper is structured as follows: in Section \ref{model_description} we describe how we model (i) the internal structure of the GMCs, and (ii) the radiation field impinging on the clouds. Then, we treat the photoevaporation theory in Sec. \ref{photoevaporation_theory} presenting the model results. In Section \ref{results_lines} we show how the evolving density field in GMCs impacts FIR line emission. We draw our conclusions in Section \ref{conclusions}, where also some caveats are discussed.

\section{Model}\label{model_description}
\begin{figure}
\centering
\includegraphics[scale=0.4]{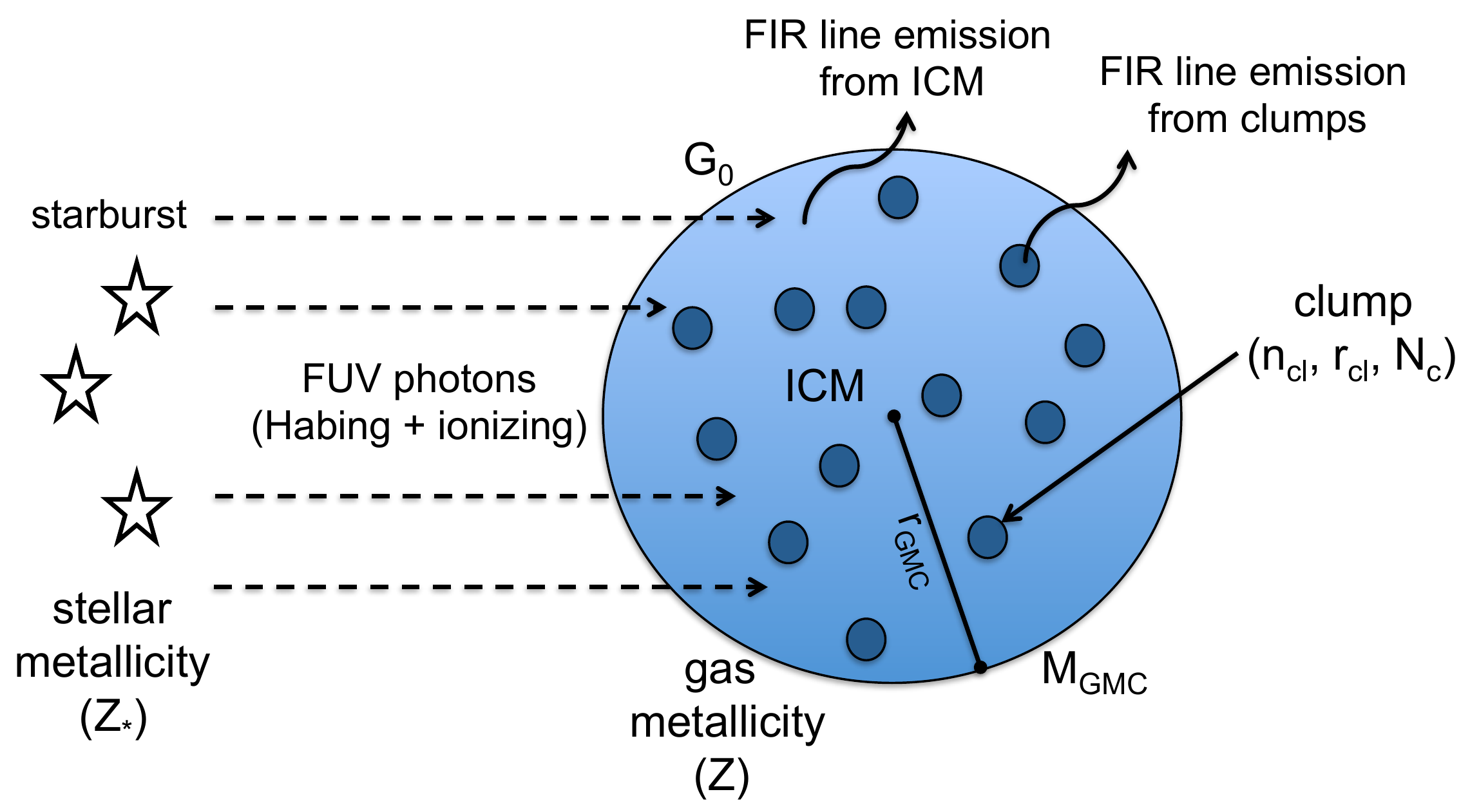}
\caption{Sketch of the GMC model used in this work.\label{sketch}}
\end{figure}

A schematic description of our model is shown in Fig. \ref{sketch}. We consider an idealized case of a clumpy, starless GMC immersed in an external radiation field. The effects of gas clumpiness on the PDR structure and the resulting FIR line emission have been studied also by \citet{meixner1993}. Those calculations, however, do not account for time evolution of the density field and feedback effects. 
The key features entering our model are: (i) the GMC density structure, and (ii) the radiation field. These are discussed in detail in the following Sections. 
\subsection{GMC density structure}\label{clump_sampling}
\begin{figure*}
\centering
\includegraphics[scale=0.7]{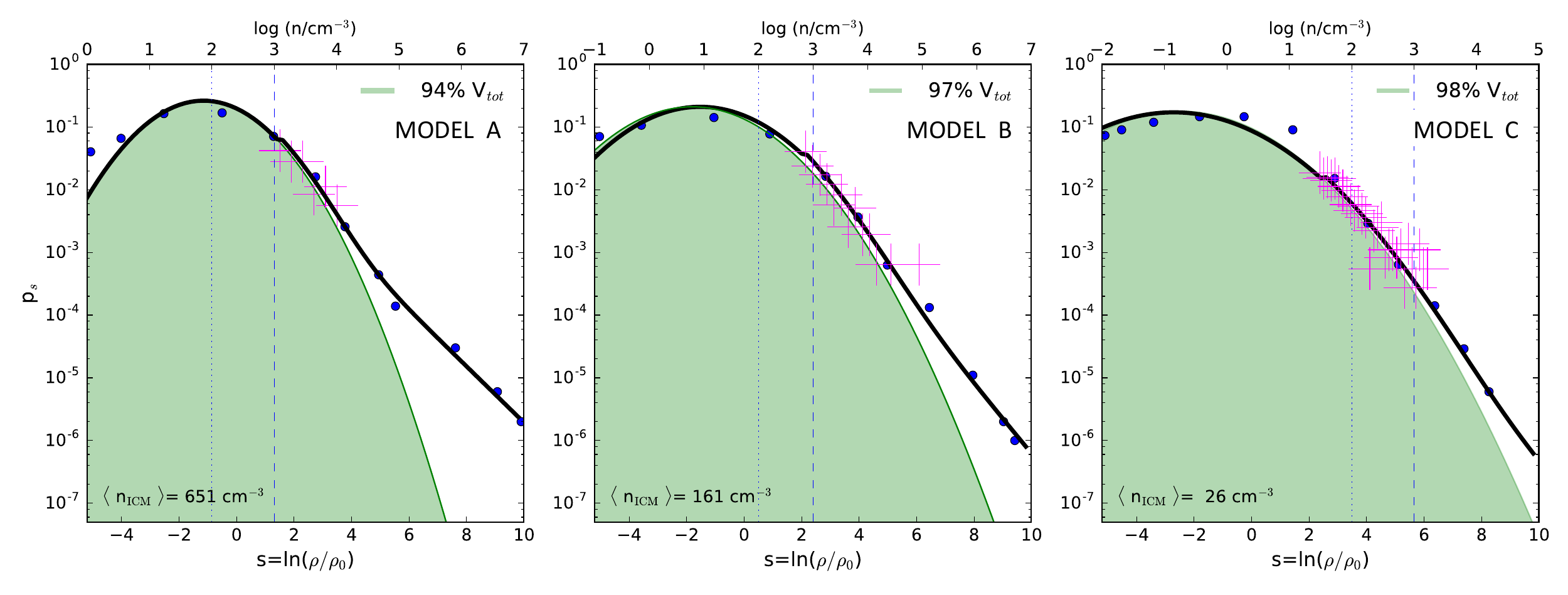}
\caption{Volume-weighted PDFs (blue points) of the logarithmic density $s\equiv \rm ln(\rho/\rho_0)$ simulated by \citet{federrath2013} for GMCs models A, B, C in Table \ref{3models}. The solid black line is the best lognormal + power-law fit to the numerical results. The volume fraction filled by the log-normally distributed gas (green area)  is given for each model. The upper $x$-axis provides the gas number density $n$, along with the reference values $n=10^2,\,10^3\,\rm{cm^{-3}}$. Magenta crosses show the density of sampled GMC clumps (see Sec. \ref{clump_sampling} for details on the procedure adopted). \label{clump_sampling2}}
\end{figure*}

Molecular clouds are observed to have a hierarchical structure with a density field showing enhancements (usually referred to as clumps and filaments) on $\approx 0.1-10$ pc scales. The typical hydrogen column density of a GMC is $N_H\approx 10^{22}\,\,\rm{cm^{-2}}$ \citep[e.g.][and references therein]{mckee2007} but variations are observed in the range $N_H\approx10^{21}-10^{23}\,\,\rm{cm^{-2}}$. GMCs are supported against collapse by turbulence and magnetic fields. 

Numerical and analytical studies conclude that the Probability Distribution Function (PDF) of the gas density, $\rho$, in a supersonically turbulent, isothermal cloud of mean density $\rho_0$ is lognormal: 
\begin{equation}\label{lognormalfunction}
g_s ds = \frac{1}{(2 \pi \sigma_s^2)^{1/2}}\, {\rm exp} \left[ -\frac{1}{2} \left( \frac{s - s_0}{\sigma_s} \right)^2 \right]
\end{equation}
with $s\equiv \rm{ln}(\rho/\rho_0)$ \citep[e.g][]{vazquez-semadeni1994, krumholz2005, padoan2011, hennebelle2011, hennebelle2013, kim2003, wada2008, tasker2009, federrath2013}. The mean logarithmic density ($s_0$) is related to the standard deviation of the distribution ($\sigma_s$) by $s_0 = -\sigma^2_s/2$ which, in turn, depends on the sonic Mach number ($\mathcal{M}$), and the ratio of thermal to magnetic pressure ($\beta$) as:
 \begin{equation}\label{lognormaleq}
\sigma^2_{s}={\ln}\left( 1+ b^2 \mathcal{M}^2 \frac{\beta}{\beta+1}\right).
\end{equation}
The $b$ factor in the previous equation parametrizes the kinetic energy injection mechanism (often referred to as forcing) driving the turbulence \citep[$b\approx0.3-1$, see][for an extensive discussion]{molina2012}. 

When self-gravity is included, the PDF develops a power-law tail ($p_s\propto \rho^{-\kappa}$) at high densities. 
The occurrence of the power-law tail is confirmed both theoretically \citep[e.g.][]{krumholz2005, hennebelle2011, padoan2011, federrath2013}, and observationally via dust extinction measurements \citep[e.g.][]{kainulainen2009, lombardi2015,stutz2015, schneider2015} or molecular line detections \citep[e.g.][]{goldsmith2008, goodman2009, schneider2015} carried out in nearby GMCs. While dust extinction allows to probe a larger dynamic range (a measured visual extinction $A_V=1-100\,\,\rm mag$ corresponds to clump column densities ($N_{cl}$) in the range $N_{cl}\approx 10^{21}-10^{23}\,\rm{cm^{-2}}$), molecular lines detections are limited to the high density tail of the PDF ($N_{cl}>10^{23}\,\rm{cm^{-2}}$). Here we consider three different cloud models (named A, B, and C, with properties summarized in Tab. \ref{3models}) in order to bracket the range of values observed in Galactic GMCs, and those assumed in simulations by \citet{federrath2013}

\begin{table}
\centering
\caption{Properties of the GMC models considered in this work \label{3models}}
\begin{tabular}{@{}cccccl@{}}
\toprule
model & $M_{\rm GMC} \,[M_{\rm \odot}]$ & $r_{\rm GMC}$ [pc] & $\mathcal{M}$ & $\rho_0$ [g cm$^{-3}$] & \\
\midrule
A     & $6.2 \times 10^3$           & $4$                  & $10$        & $8.2 \times 10^{-22}$ &  \\
B     & $9.9 \times 10^4$           & $16$                 & $20$        & $2.1 \times 10^{-22}$ & \\
C     & $3.9 \times 10^6$           & $100$                & $50$        & $3.3 \times 10^{-23}$ & \\
\bottomrule
\end{tabular}
\end{table}

We set up the internal density of the clouds so that their PDFs are in agreement with that found by \citet{federrath2013} for $\mathcal{M}= 10,\, 20,\, 50$, mean gas density $\rho_0=8.2 \times 10^{-22},\, 2.1 \times 10^{-22},\,3.3\times 10^{-23} \,\rm{g \, cm^{-3}}$, and turbulence forcing parameter $ b\approx0.3$. In Fig. \ref{clump_sampling2}, we show the PDF that has been fitted with a lognormal ($g_s$, eq. \ref{lognormalfunction}) function + power-law ($t_s$) tail, i.e. $p_s = g_s+t_s$. In what follows, we identify two components in the GMC, referred to as \textit{clumps} and diffuse \textit{interclump medium} (ICM), adopting a criterion based on the density PDF. We use the term \textit{clumps} to denote small scale structures ($< 1\,\rm{pc}$ in size) that are part of the power-law tail. In the literature, clumps have been interpreted either as temporary density fluctuations produced by supersonic turbulence \citep[][]{falgarone1990}, or as stable physical entities confined by ICM pressure \citep{williams1995}. Even though clump morphology is observed to vary from filamentary to quasi-spherical shapes, in our work we model the clumps as spheres (see Fig. \ref{sketch}).

The volume filled by the gravitationally unstable clumps in the GMCs is:
\begin{equation}
V_{clumps}=V_{\rm GMC} -  V_{\rm ICM}
\end{equation} 
where the volume of the ICM is given by the log-normal distribution
\begin{equation}
V_{\rm ICM} = V_{\rm GMC} \int_{} g_s \,ds.
\end{equation} 
We obtain $\int_{} g_s \,ds=0.94,\,0.97,\, 0.98\, $ for models A, B, and C, respectively (see Fig. \ref{clump_sampling2}).
To build the internal density field we then use the following procedure:
\begin{itemize}
\item[1.] Randomly extract from the tail $t_s$ the i-th clump with number density\footnote{We assume the gas to be a mixture of hydrogen and helium with mean molecular weight $\mu =1.22$.} $n_{cl,i}=\rho/(\mu m_p)$.
\item[2.] Calculate the clump radius as the turbulent Jeans length:
\begin{equation}
r_{cl,i}=\frac{1}{2} \lambda_{J, turb}= \frac{1}{2} \frac{\pi \sigma^2 + \sqrt{36 \pi c_s^2 G L^2 \rho + \pi^2 \sigma^4}}{6GL\rho}\label{jeans_fk}
\end{equation}
where $c_s$ is the sound speed. Eq. \ref{jeans_fk} is obtained by using an effective (turbulent+thermal) pressure term in the Jeans length equation \citep[see eqs. 35, 36 in][]{federrath2012}.
\item[3.] Calculate the clump volume $V_{cl,i}= (4/3 )\pi r_{cl,i}^3$.
\item[4.] Iterate steps 1-3 until $\Sigma_{i} V_{cl,i} = V_{clumps}$.
\end{itemize}
Finally, we compute the total mass in clumps $M_{tot, cl}=\Sigma_{i}\frac{4}{3} \pi \mu m_{p} n_{cl,i} r_{cl,i}^3$, the ICM total mass $M_{\rm ICM}=M_{GMC}- M_{tot, cl}$, and the ICM mean density $\langle n_{\rm ICM} \rangle=M_{\rm ICM}/(\mu m_p V_{\rm ICM})$.
The resulting clump distributions are shown with magenta crosses in Fig. \ref{clump_sampling2}. For model A, B and C we find $M_{tot, cl}\approx 1.2 \times 10^{3},\, 1.8 \times 10^{4},\,7.2 \times 10^{5}\,\rm{M_{\odot}}$, and $\langle n_{\rm ICM}\rangle\approx 651,\, 161,\, 26 \,\rm{cm^{-3}}$, respectively.

\subsection{Radiation field}\label{starburst_spectrum}

\begin{figure}
\centering
\includegraphics[scale=0.4]{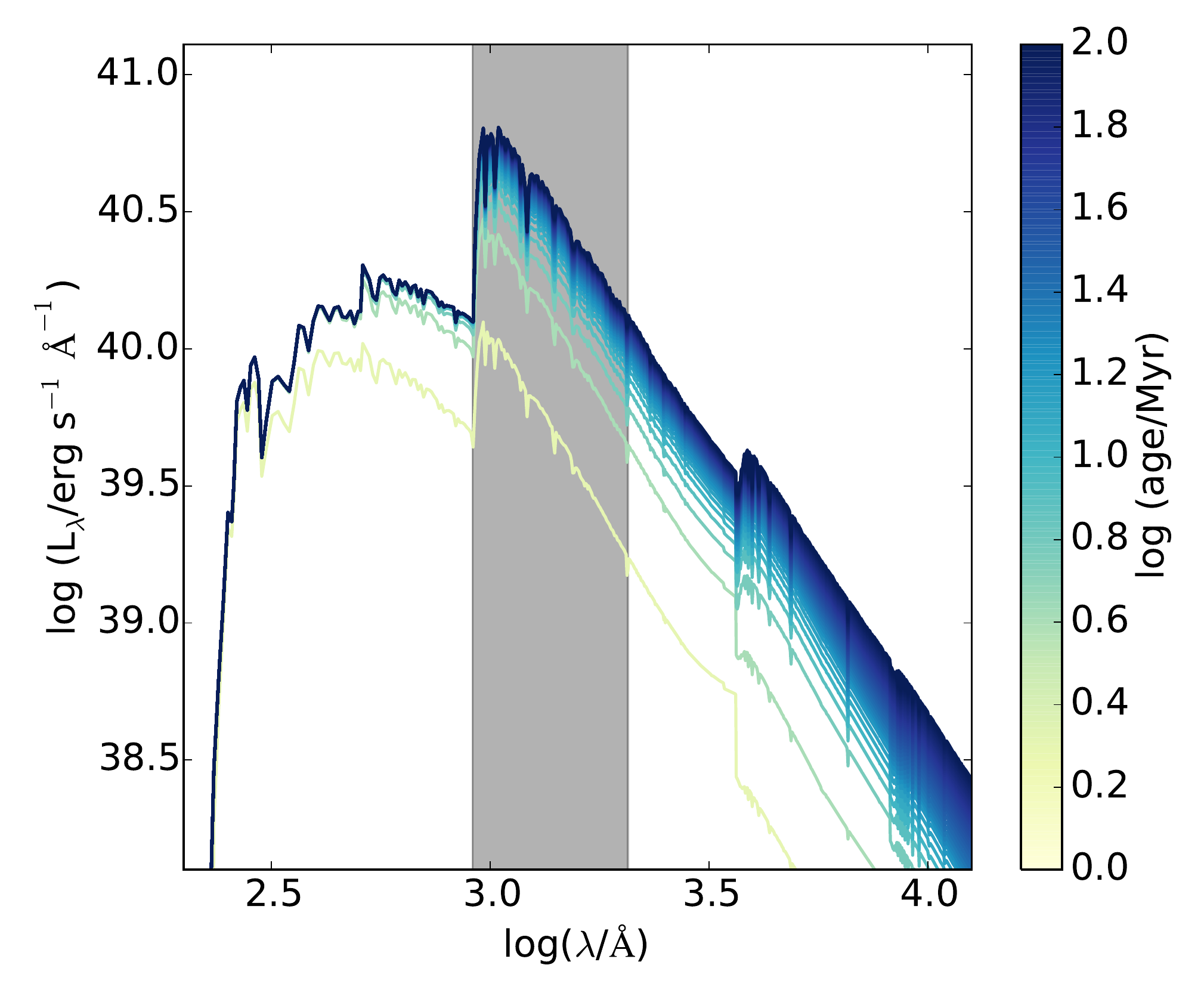}
\caption{Spectral energy distribution (SED) of the radiation field produced by stars with $Z_*=0.2\,Z_{\odot}$ as a function of time elapsed from the onset of the star formation. We assume a continuous SFR of $1\,\rm{M_{\odot}\,yr^{-1}}$. The gray shaded region highlights the Habing band.}\label{sb9}
\end{figure}
The spectral energy distribution (SED) of the radiation field impinging on the GMC surface is calculated using the stellar population synthesis code {\tt Starburst99} \citep{leitherer1999}, assuming a Salpeter Initial Mass Function in the range $1-100\,\rm M_\odot$.
We adopt the Geneva standard evolutionary tracks \citep{schaller1992} with metallicity $Z_*=\rm{1\, Z_{\odot}} ,\, 0.2 \,Z_{\odot}, \rm{\,and\,} 0.05 \,\rm{Z_{\odot}}$, and Lejeune-Schmutz stellar atmospheres which incorporate plane-parallel atmospheres and stars with strong winds \citep{lejeune1997, schmutz1992}.
We follow the time evolution of the SED between {$1-100$ Myr} considering a continuous star formation mode. The star formation rate is a free parameter of the model. As an example, we show the SED for SFR $=1\,\rm{M_{\odot}\,yr^{-1}}$ and $Z=0.2\,\rm{Z_{\odot}}$ in Fig. \ref{sb9}.
The lines are color-coded as a function of the starburst age. The gray shaded region highlights the non-ionizing FUV Habing band relevant to the PDR modeling; the strength of the FUV radiation is usually parameterized by $G_0$, the ratio of the FUV flux to the one measured by \citet{habing1968} in the Milky Way ($\approx 1.6\times 10^{-3}\rm {erg\, cm^{-2}\, s^{-1}}$). As expected for a continuous star formation mode, the specific luminosity in the Habing band increases with time before saturating to an asymptotic value around 100 Myr\footnote{The SED can be approximated as a power-law of the form $\log L_{\lambda}= \alpha\log \lambda +{\rm const}$ for $\lambda>912$ \AA. At the times relevant for photoevaporation, we find $\alpha= -2.2$ (age 1 Myr) and $\alpha=-2.4$ (age 10 Myr).}.
%

%
\section{Photoevaporation}\label{photoevaporation_theory}
The UV radiation produced by massive OB stars influences the structure, dynamics, chemistry, and thermal balance of the surrounding gas. 
Ultraviolet photons substantially alter the clump-interclump structure of GMCs: EUV and/or FUV photons heat the surface layer of clumps to high temperatures causing the loss of their cold molecular mass that is ionized and/or photodissociated and it is converted into warm ionized/atomic gas \citep{hollenbach1999}. This process is called \textit{photoevaporation}. When the GMC is embedded in an HII region, the ICM is exposed to both EUV and FUV photons, while the internal clumps see only the attenuated FUV radiation. Penetration of EUV and FUV photons in the ICM and clumps is further addressed in Appendix  \ref{cloudy_simulations}. There we show the temperature, $x_{\rm HI}$, and $G_0$ profiles obtained with photoionizaton simulations discussed in detail in Sec. \ref{param_eval}.

\begin{figure}
\centering
\includegraphics[scale=0.4]{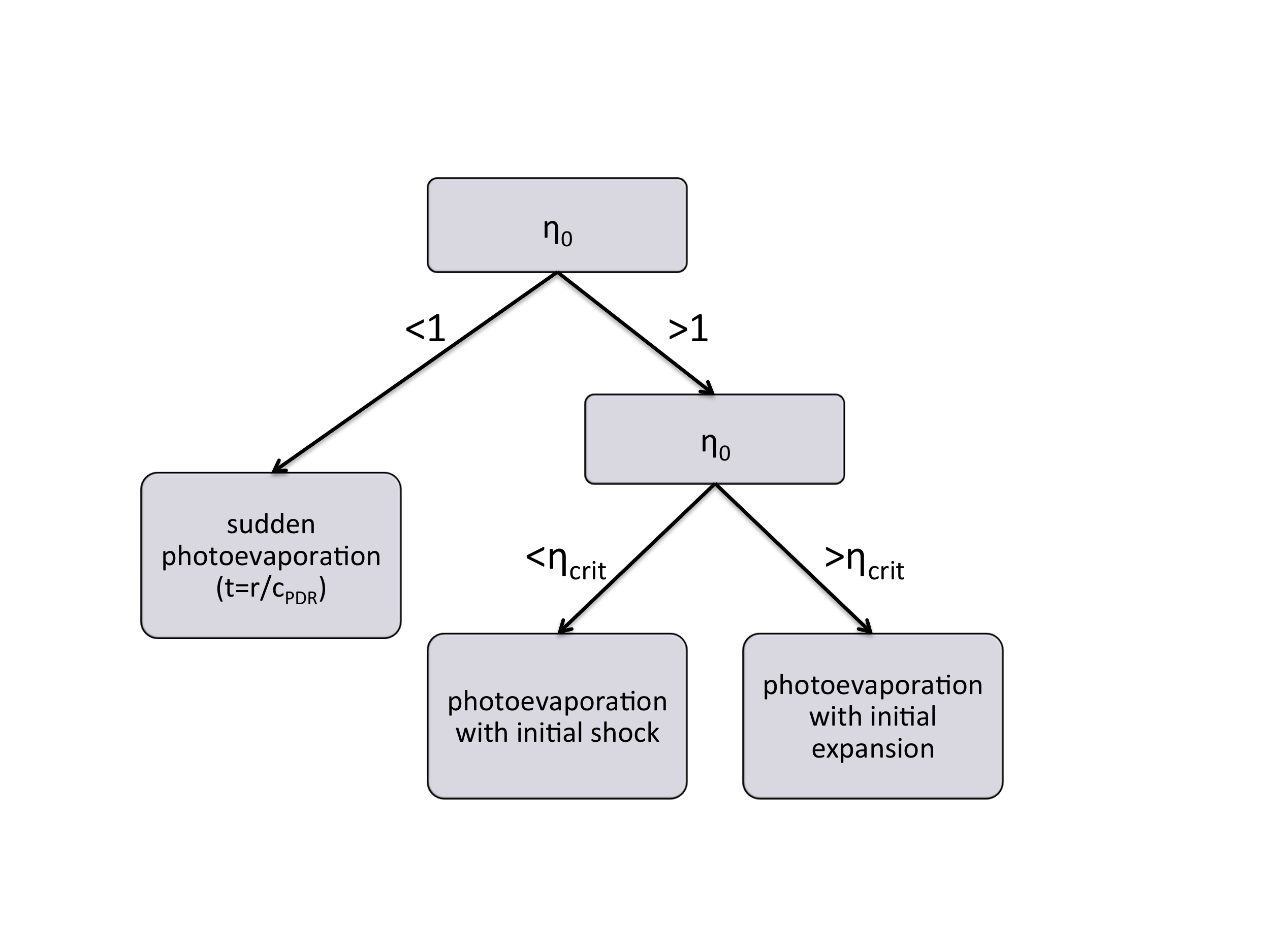}
\caption{The three channels of the photoevaporation process. See text for the discussion of the various bifurcation parameters.} \label{photoev_chan}
\end{figure}
\begin{figure*}
\centering
\includegraphics[scale=0.45]{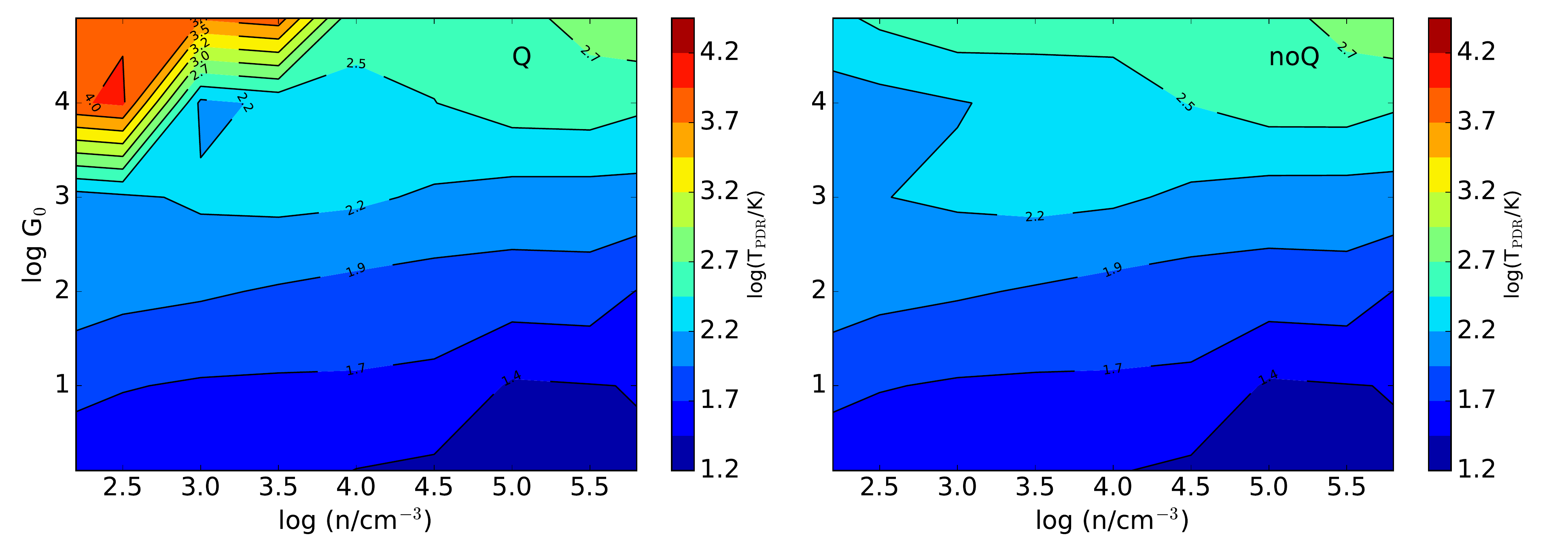}
\caption{PDR temperature as a function of the gas density, $n$, and Habing flux, $G_0$, for {\tt Q models} (left) and \texttt{noQ models} (right) for a solar metallicity gas. $T_{\rm PDR}$ is measured where the FUV ($\lambda_{ref} =1000$\AA) optical depth $\tau_{\rm FUV} = 1$.} \label{pdr_properties}
\end{figure*} 

\subsection{Analytical approach}

\citet{gorti2002} show that the evolution of a turbulent clump, impulsively irradiated by FUV photons is influenced only by two parameters: \textit{(a)} the ratio of the clump initial column density ($n^0_{cl} r^0_{cl}$) to the column ($N_0$) penetrated by the FUV radiation, 
\begin{equation}\label{first_param}
\eta_{0}\equiv \frac{n^0_{cl} r^0_{cl}}{N_{0}},
\end{equation}
i.e.  the depth into the cloud where $\tau_{FUV}\approx1$; \textit{(b)} the strength of the FUV field, parametrized by the ratio of the sound speed in the FUV-heated region ($c_{\rm PDR}$) to the sound speed of the clump in the no-field case\footnote{The parameter $\nu$ is therefore proportional to the square root of the ratio of the gas temperature in the PDR ($T_{\rm PDR}$), over the initial temperature of the clump ($T_c$)}:
\begin{equation}\label{second_param}
\nu \equiv c_{\rm PDR}/c_{c}.
\end{equation}
If the ICM, as in the turbulent origin scenario previously discussed, has negligible pressure compared to the clumps, their evolution depends on the value of $\eta_0$. Small clumps with $\eta_{0}\leq 1$ are rapidly (on timescales of order of $t_c\approx r_{cl}/c_{\rm PDR}$) heated and photodissociated by the FUV flux. Larger clumps ($\eta_{0}>1$) can confine the FUV-heated region to a thin surface layer. For large clumps, an additional bifurcation point exists (see the sketch in Fig. \ref{photoev_chan}) at a critical value\footnote{The critical value can be derived from mass conservation and from the condition of pressure equilibrium at the surface of the clump. For details see \citet{gorti2002}.} $\eta_{crit}\approx 4\nu^2/3$.  

\begin{itemize}
\item If $\eta_{0}<\eta_{crit}$ the FUV radiation produces a PDR shell whose pressure is initially much higher than that of the cold clump gas and a shock is driven into the clump, compressing it. $N_{cl}$ remains roughly constant due to mass loss via photoevaporation; as a result the density increases during the compression phase. Once the shock has reached the center, the shrinking rate slows down, and it is purely regulated by mass loss from the surface.

\item If $\eta_{0}>\eta_{crit}$, the PDR shell is very thin compared to the clump size. The outer edge of the shell expands at roughly $c_{\rm PDR}$, and its pressure drops rapidly. Hence, the clump pressure eventually becomes higher than the shell one triggering the expansion also of the clump gas (at the corresponding sound speed $c_c$). This expansion continues until pressure balance with the outer PDR layers is established. At this point the cold clump loses mass gradually, with an evolution similar to the final solution for shock-compressed clumps.
\end{itemize}
\citet{gorti2002} derive analytical equations describing the time evolution of the density, radius and mass of the clumps in the two regimes. We refer the reader to Appendix \ref{appendices} for details, and to \citet{gorti2002} for a complete discussion.
\subsection{Parameter evaluation}\label{param_eval}
The parameters $\eta_0$ and $\nu$ depend on the PDR column density $N_0$ (see Eq. \ref{first_param}) and temperature $T_{\rm PDR}$ (see Eq. \ref{second_param}). In turn, these quantities depend on the clump density, $n_{cl}$, and $G_0$:
\begin{eqnarray}
N_0 =& N_0 (G_0, n_{cl});\,\,\,\,\, T_{\rm PDR} = & T_{\rm PDR} (G_0, n_{cl}).
\end{eqnarray}\label{cloud_procedure}

We calculate $N_0$ and $T_{\rm PDR}$ with version c13.03 of \cloudy \, \citep{ferland2013}, which allows us to model the transition between the \HII region, PDR, and molecular part of a gas slab illuminated by a given radiation field. 
For each of the three metallicities\footnote{We do not make a distinction between gas and stellar metallicities, which are then supposed to be equal} considered in this work ($Z=0.05,\,0.2,\,1 \,\rm Z_{\odot}$) we run two sets of simulations with different prescriptions for the external radiation spectrum: (a) a full spectrum (\texttt{Q models}) including both FUV and EUV photons, (b) a FUV spectrum only (\texttt{noQ models}). The \texttt{Q models} (\texttt{noQ models}) are designed to mimick the flux reaching GMC placed within (outside) an \HII region.

We run a total of $66\times3$ \cloudy \, simulations for each metallicity, and for ${\rm log}(n/\rm{cm^{-3}})=[1-6]\,$ (in steps of 0.5 dex), and ${ \rm log} \,G_0=[0 - 5]$ (1 dex). The parameter space covers the plausible range of clumps/ICM densities (see Sec. \ref{clump_sampling}), and Habing fluxes in galaxies. The code computes the radiative transfer through the slab up to a hydrogen column density $N_{\rm H}=10^{23}\,\rm{cm^{-2}}$. This stopping criterium is chosen to \textit{(i)} cover the whole range of column densities of our randomly generated clumps, and \textit{(ii)} to fully sample the molecular part of the illuminated slab, typically located at $N_{\rm H}\simgt 2\times10^{22}\,\rm{cm^{-2}}$. 

We adopt the gas-phase abundances ($\rm C/H=3.0 \times 10^{-4}$, $\rm O/H=4.0 \times 10^{-4}$, $\rm Mg/H=3.0 \times 10^{-6}$, $\rm N/H=7.0 \times 10^{-5}$, $\rm S/H =1 \times 10^{-5}$) provided by \cloudy \, for the Orion Nebula \citep{rubin1991, osterbrock1992, rubin1993}\footnote{
As a caveat, we note that, in the standard \cloudy~ set for the Orion Nebula, the carbon and oxygen abundances provided are $\approx 2$ and $\approx 1.5$  times greater than the values reported by e.g. \citet[][$\rm C/H=1.4 \times 10^{-4}$]{cardelli1996} and \citet[][$\rm O/H=2.8 \times 10^{-4}$]{cartledge2004}, respectively.}, scaled with the metallicity of each specific model. The model accounts for the CMB background at $z=6$, that can suppress the emergent line luminosity of FIR lines when observed in contrast with the CMB \citep{gong2012, dacunha2013, vallini2015, pallottini2015, zhang2016}
In the calculation we consider a cosmic-ray (CR) ionisation rate $\zeta_{\rm CR} =2 \times 10^{-16} \rm s^{-1}$ \citep{indriolo2007}. 
Note, as a caveat, that the variation of the CR ionisation rate has strong effects on the chemistry and emission of PDRs \citep[e.g.][]{papadopoulos2011, bayet2011, meijerink2011, bisbas2015}.\footnote{The effect of the variation of $\zeta_{\rm CR}$ on the [\CII] (CO) line intensity has been quantified by e.g. \citet{meijerink2011} who show that for a PDR of density $n=10^3\,\rm{cm^{-3}}$, irradiated by $G_0=10^3$, $I_{\rm CII}= [4.4-8.6] \times 10^{-4} \,\rm erg s^{-1} sr^{-1}$ ($I_{\rm CO}=[ 7.6- 0.28 ] \times 10^{-8} \,\rm erg s^{-1} sr^{-1}$) when varying $\zeta_{\rm CR} =5 \times [10^{-17}- 10^{-14}] \, \rm s^{-1}$. Recently \citet{bisbas2015} have argued that CR-induced destruction of CO in GMCs is likely the single most important factor controlling the CO-visibility in star-forming galaxies.}  The CRs, unlike FUV radiation, travel nearly unimpeded through the clouds and provide a source of input energy by: (i) freeing electrons, and (ii) inducing an internal UV field through the excitation of $\rm H_2$  \citep[see also][and references therein]{indriolo2013}.

In Figure \ref{pdr_properties}, we plot $T_{\rm PDR}$ as a function of the gas number density ($n$) and Habing flux ($G_0$) for $Z=Z_\odot$ where the FUV ($\lambda_{ref} =1000$\AA) optical depth $\tau_{\rm FUV} = 1$. This criterium is chosen because the temperature $T_{\rm PDR}$ entering in the \citet{gorti2002} model (see Sec. \ref{photoevaporation_theory}), refers to the FUV-heated region, i.e. extending up to the point at which the FUV optical depth reaches a value of the order of unity.

The PDR temperatures obtained with \texttt{Q models} and \texttt{noQ models} are similar, apart from low-density ($n<10^{3}\,\rm{cm^{-3}}$ and strong field ($G_0>10^3$) regimes, where \texttt{Q models} are warmer. The results are in agreement with those found by \citet{kaufman1999} \citep[see also Fig. 19 in][]{visser2012}.  Furthermore, as a sanity check, we compare $T_{\rm PDR}$ with that obtained by \citet{bothwell2016} using the \textsc{3-d pdr} code \citep{bisbas2012}. \citet{bothwell2016} cover the same range of $G_0$ and $n$ considered here, but they sample the PDR temperature deeper into the gas slab ($\rm A_V=3$). As expected, our results ($T_{\rm PDR}\approx 10^{1.2-2.5} \rm K$, in the range $G_0=[10^0-10^4]$ and $n=[10^2-10^5]\, \rm {cm^{-3}}$) are slightly higher than those ($T_{\rm PDR}\approx 10^{1.1-2.0} \rm K$) found by \citet{bothwell2016} (see their Fig. 10) in the same range of $n$ and $G_0$.

Once the PDR temperature and column density as a function of $n$ and $G_0$ are known, we can determine the values of $\eta_0$ and $\nu$ for each cloud. This allows us to compute the time evolution of their mass, density and radius (see Appendix \ref{appendices}). At each time step $t_i$, we update the value of $\eta_0$ and $\nu$ according to the clump density at the previous time step, $n_{cl}(t_{i-1})$, assuming that $G_0$ is constant with time. 
\citep{bothwell2016, visser2012}
\subsection{Clump photoevaporation}

\begin{figure*}
\centering
\includegraphics[scale=0.85]{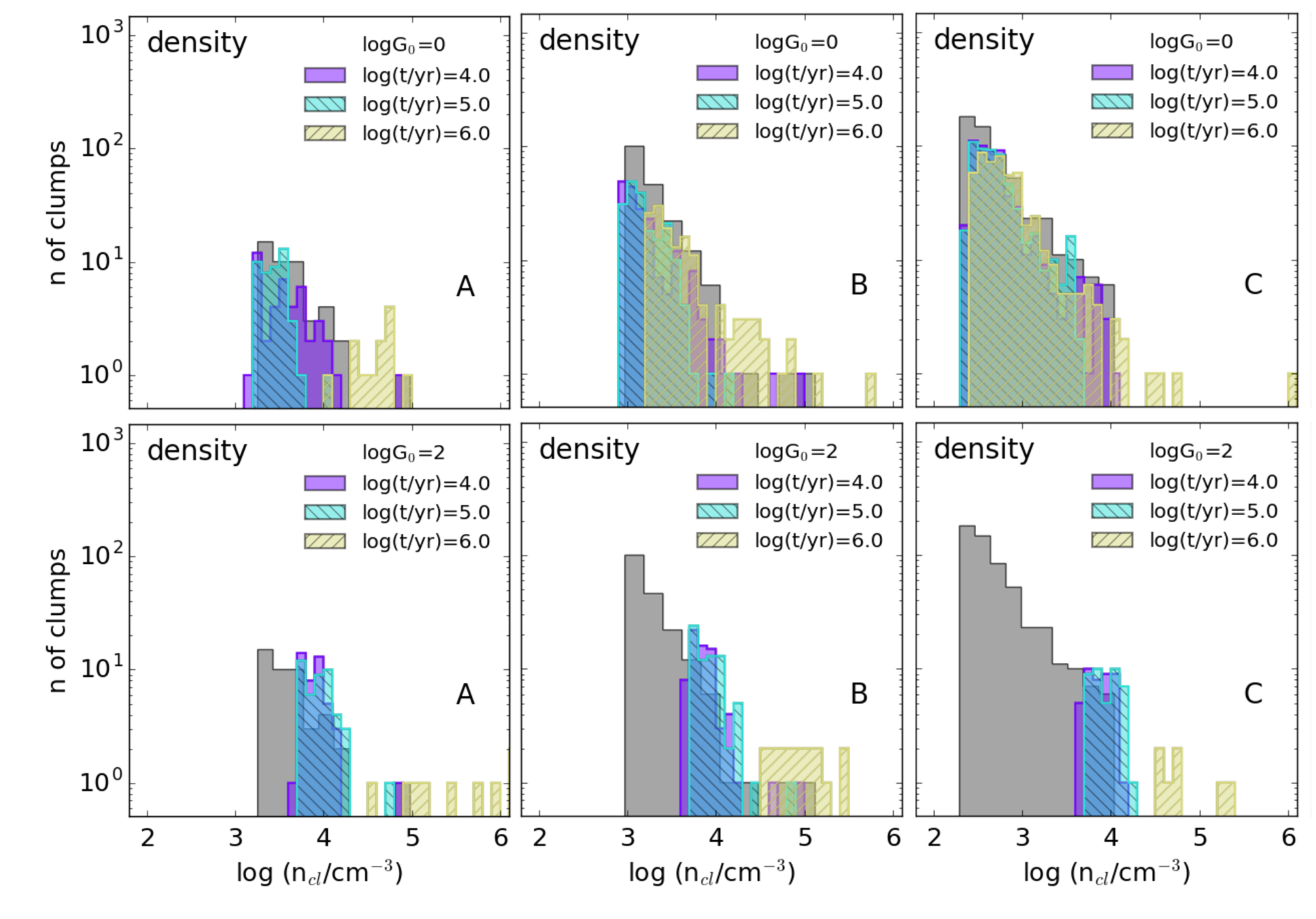}
\caption{Time evolution of the clump density distributions for models A, B, C as a function of $G_0$ at the clump surfaces ($\rm log G_0 =0, \,2$, top and bottom row respectively). The distributions at $t=10^4,\, 10^5,\,10^6\, \rm{yr}$ are shown with transparent colored histograms, and the initial distribution is shown in solid gray. Photoevaporated clumps are removed from the distribution. \label{clump_evol}}
\end{figure*}

\begin{figure*}
\centering
\includegraphics[scale=0.85]{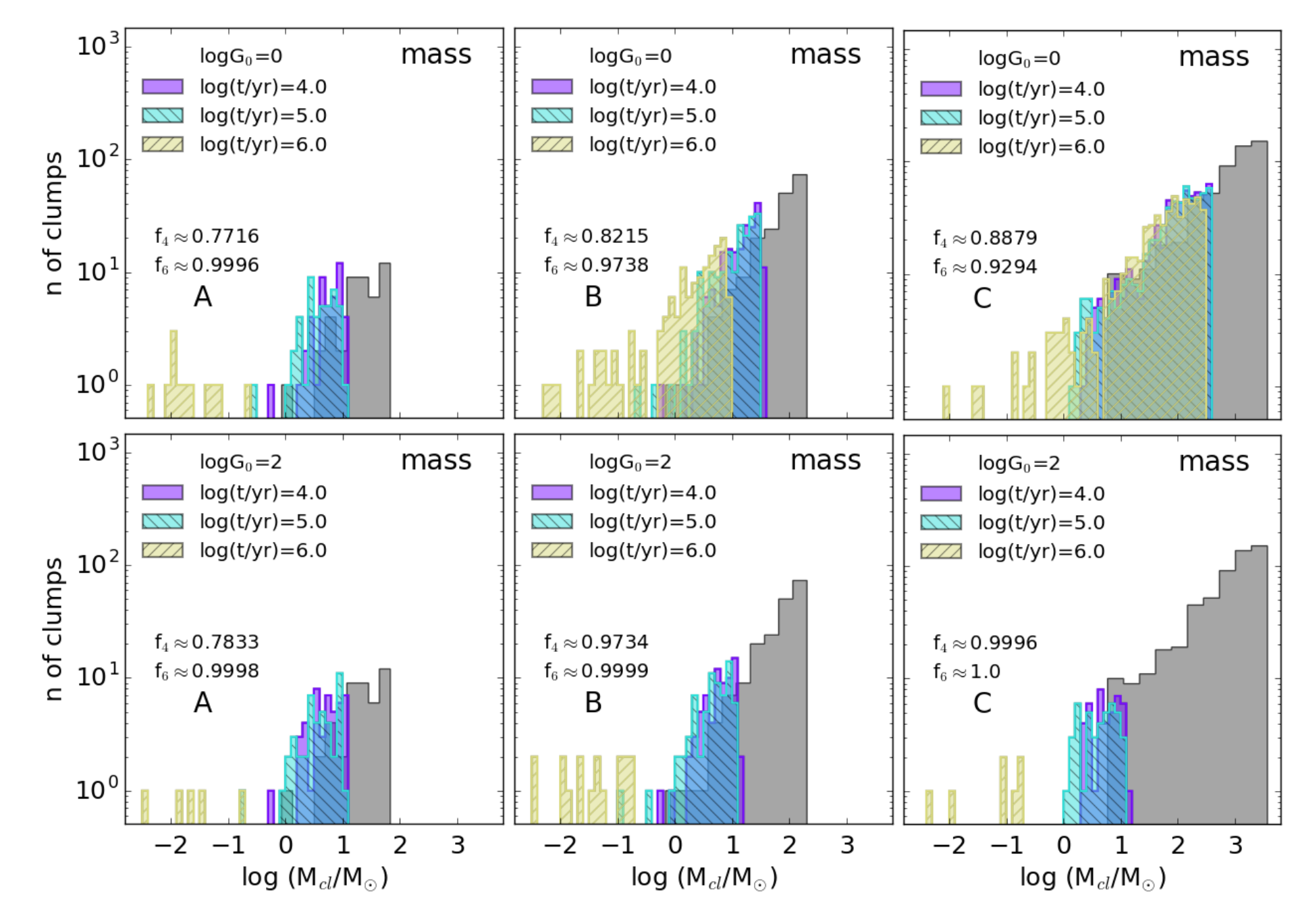}
\caption{Same as Fig. \ref{clump_evol} but for the clump mass distribution. The fraction of mass returned to the ICM at $t=10^4,\,10^6\, \rm{yr}$ ($f_4,\,f_6$, respectively) is given in each panel. \label{clump_evol2}}
\end{figure*}

Figs. \ref{clump_evol} and \ref{clump_evol2} show  the time evolution of the clump density and mass distribution for the three model clouds A, B, C. We concentrate on the effect $G_0$ variations on clumps photoevaporation at fixed metallicity, $Z=\rm{Z_{\odot}}$. We define a clump as completely photoevaporated at time $t$, and hence removed from the clump inventory, if one of the two following conditions is satisfied: (i) the radius $r_{cl}(t)=0$, or (ii) the clump density $n_{cl}(t)=\langle n_{\rm ICM} \rangle$, i.e. the clumps become indistinguishable from the ICM.
%
For ${\rm log} \,G_0=0$ the photoevaporation proceeds mainly via expansion. The density distribution shifts toward lower values. However, a small fraction ($<1\%$ in mass, see Fig. \ref{clump_evol2}) of longer-lived, compressed clumps is visible in the high density tail at $t=10^6\,\rm yr$. On the contrary, for ${\rm log} \,G_0=2$, photoevaporation proceeds via shock compression for all the clumps, and the distribution shifts toward higher densities. These trends hold for both model A and B;  model C is initially less dense (see Fig. \ref{clump_sampling2}) and the compression mode becomes important already for ${\rm log} G_0 =0$. 

During photoevaporation mass loss takes place, and a certain fraction of the clump mass is returned to the ICM. For $\log\,G_0 = {0}$, such fraction after $10^4$ yr is $f_4=0.77-0.88$ depending on the cloud model; for $\log\,G_0 = 2$ mass loss is more substantial, $f_4=0.78-0.99$. 
We note that fixed $G_0$ at the clump surface, $f_4$ and $f_6$ (the analogous fraction after 1 Myr) increase going from model A, to B, to C, along their decreasing clump density sequence.

\subsection{ICM photoevaporation}

ICM photoevaporation is computed with a procedure similar to that adopted for the clumps. 
However, for the ICM the effects of EUV photons become important if the GMC is located within an \HII region. 
For example, consider the case $\rm log\,G_0 \geq5$ and $Z=Z_{\odot}$. Then, the typical column density of the \HII layer in a gas of $n\approx 100\, \rm{cm^{-3}}$ is $N_{\rm HII}\approx10^{22}\,\rm{cm^{-2}}$ (see Fig. \ref{profile_ionization}, and the discussion in Appendix \ref{cloudy_simulations}), and thus comparable to the ICM column density. This implies that the GMC is almost fully ionized, and that photoevaporation is driven by the increased temperature ($T_{\rm HII} \approx 10^4\,\rm{K}$) in the ionized layer.

To model this regime we adopt a modified version of the photoevaporation equations in Appendix \ref{appendices}. In the equations for the time evolution of radius, mass, and density, we replace the parameter $\eta_0$ (see Eq. \ref{first_param}) with:
\begin{equation}\label{first_param_2}
\eta_0^{\rm HII} \equiv \frac{n_{\rm ICM}\,r_{\rm GMC}}{N_{\rm HII}},
\end{equation}
where $N_{\rm HII}$ is the column density of the \HII layer. Moreover, we substitute $\nu$ (see Eq. \ref{second_param}) with:
\begin{equation}\label{second_param_2}
\nu^{\rm HII} \equiv \frac{c_{\rm HII}}{c_c}
\end{equation}
where $c_{\rm HII}$ is the sound speed in the ionized layer.
Note that $\nu_{\rm HII}>\nu$, due to the higher temperature in the \HII region with respect to PDRs. 
As in the case of clumps, $\eta_0^{\rm HII}$ and $\nu_{\rm HII}$ depend on $N_{\rm HII}$ and $T_{\rm HII}$, which have been determined from \cloudy \, simulations at the depth at which the gas is 50\% ionized.  
We follow the evolution of ICM density, GMC radius and mass until complete GMC photoevaporation, defined by one of the two criteria: (i) $M_{\rm ICM}<10\% \, M_{\rm GMC}$, or (ii) the ICM density falls below $\approx10\,\rm{cm^{-3}}$, i.e. the typical density of the diffuse cold neutral ISM phase \citep{wolfire2003}. 

\subsection{GMC photoevaporation timescales}

\begin{figure*}
\centering
\includegraphics[scale=0.5]{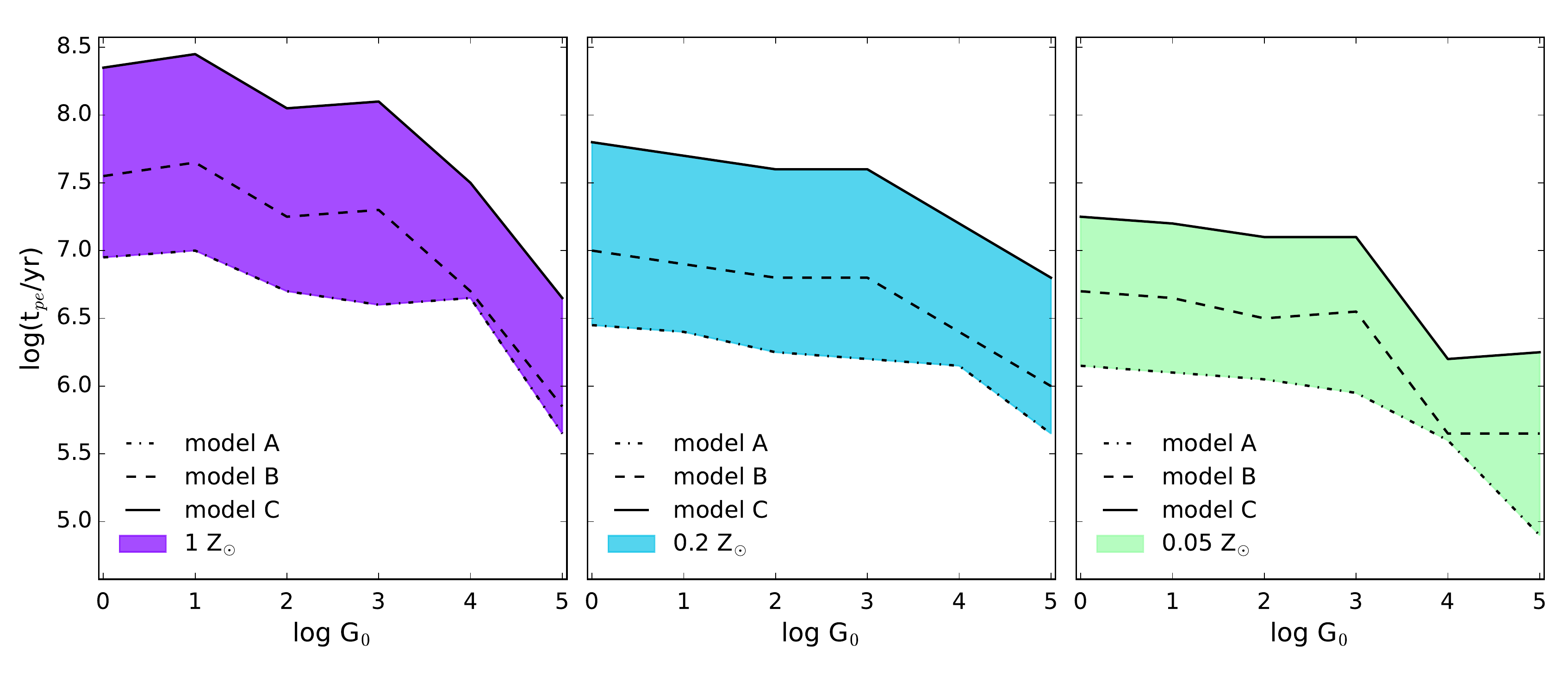}
\caption{GMC photoevaporation time (t$_{pe}$) vs. $G_0$ for $Z=1,\,0.2,\,0.05\,\rm{Z_{\odot}}$ from left to right). In each panel, the dashed line represents $t_{pe}$ for our fiducial cloud (model B). The shaded area highlights the variation of $t_{pe}$ for model C (solid lines) and model A (dot-dashed lines).}\label{fig_ICM_photoevap}
\end{figure*}

In Fig. \ref{fig_ICM_photoevap} we plot the GMC photoevaporation time ($t_{pe}$) as a function of $G_0$ for different metallicities, $Z=1,\,0.2,\,0.05\,\rm{Z_ \odot}$. At fixed cloud mass and $G_0$, a decreasing $Z$ results in a faster photoevaporation. This trend is mostly driven by the decreasing dust-to-gas ratio allowing the deeper UV radiation penetration to heat the internal gas layers.
As a final remark we note that, for $Z=Z_{\odot}$ the maximum $t_{pe} \approx 30 \,\rm{Myr}$ in model B is consistent with the results by \citet{williams1997, krumholz2006}. These authors find $t_{pe}\approx30-40 \,\rm Myr$ for GMCs of mass $10^{5}\rm{M_{\odot}}$ when considering the photoevaporation produced by OB associations \textit{inside} the GMC.

\section{FIR line emission}\label{results_lines}
\begin{figure*}
\includegraphics[scale=0.5]{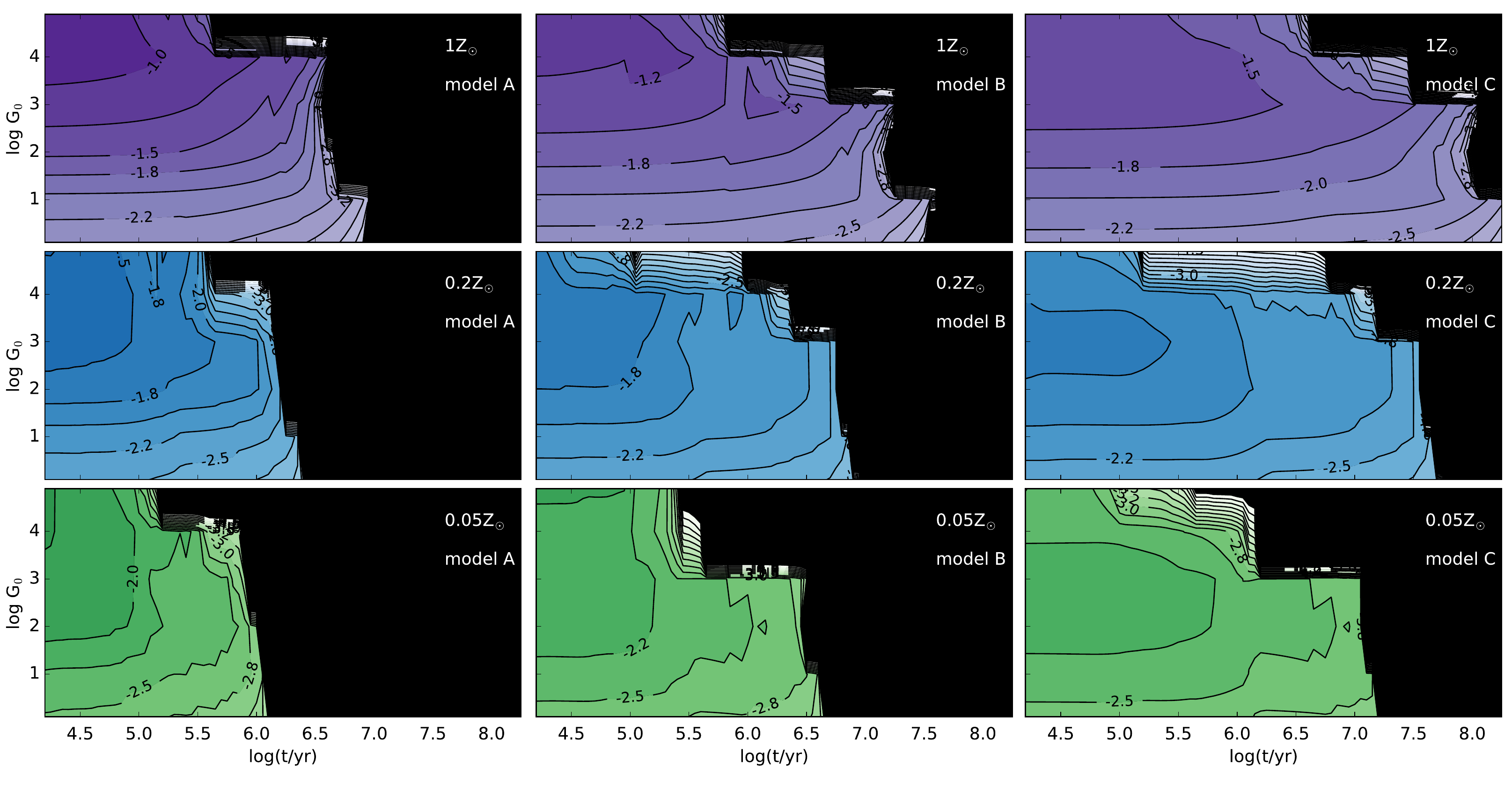}
\caption{Specific [\CII] luminosity (in units of $L_\odot/M_\odot$) evolution for cloud models A, B, C from left to right respectively as a function of $G_0$ and different metallicities ($Z=Z_{\odot}$ upper panels, $Z=0.2\,Z_{\odot}$ middle, and $Z=0.05\,Z_{\odot}$ bottom). Black regions denote $\epsilon_{CII} < 10^{-5}$.\label{specific_cii}}
\end{figure*}
From our model, we now compute the FIR line emission from GMCs, including the effects of PE. At each time step in the computation, we derive the line luminosity of the clumps (ICM) depending on their actual density, $n_{cl}(t)$ ($n_{\rm ICM}(t)$), emitting area, $\pi r_{cl}(t)^2$ ($\pi r_{\rm GMC}(t)^2$), and column density $N_{cl}(t)=n_{cl}(t)r_{cl}(t)$ ($N_{ICM}(t)=n_{\rm ICM}(t)r_{\rm GMC}(t)$). Again, we use \cloudy to compute the FIR line flux, $I_{\rm line} (n,G_0, N_H)$ (in $\rm erg\,s^{-1}\,cm^{-2}$), at the relevant surface (clump/ICM). The total luminosity is calculated as a sum over the clumps and the ICM:
\begin{equation}
\begin{aligned}
L_{\rm line} (t) = &  \sum_{cl} I_{\rm line} \left(\langle f_{\rm att} \rangle G_0, n_{cl}(t), N_{cl}(t)\right)\, \pi r_{cl}^2 (t) + \\
& I_{\rm line} \left(G_0, n_{\rm ICM}(t), N_{\rm ICM}(t)\right)\, \pi r_{\rm GMC}^2 (t).
\end{aligned}
\end{equation}
In the above expression, the UV flux seen by the clumps is attenuated by the ICM by an average factor $\langle f_{\rm att}\rangle=G_0(\langle N_{\rm H} \rangle)/G_0$ to roughly account for GMC-scale radiative transfer effects not included here.  The mean absorbing column density to each clump is $\langle N_{\rm H} \rangle \approx 0.25 \, r_{\rm GMC}\, n_{\rm ICM}$. For models A, B, C, we get $\langle N_{\rm H} \rangle \approx 2 \times 10^{21}\,\rm{cm^{-2}}$, yielding $\langle f_{\rm att} \rangle \approx 0.2,\,0.72,\, 0.9$ for $Z=1,\,0.2,\,0.05 \,{\rm Z_{\odot}}$. See App. \ref{cloudy_simulations} for further discussion on this point.

The predicted specific (i.e. per unit mass of emitting material)  [\CII] luminosity, $\epsilon_{\rm CII}$, is shown in Fig. \ref{specific_cii} for the different cloud models. 
Such predictions are in very good agreement with recent observations, e.g., of the Orion Molecular Cloud 1 (OMC1) by \citet{goicoechea2015}. The observed total mass in OMC1 region is $M_{gas}=2600\,\rm{M_{\odot}}$, thus comparable with $M_{\rm GMC}$ in model A. Additionally, \citet{goicoechea2015} measured a mean value of $G_0\simeq 2 \times 10^4$, and a specific luminosity $L_{\rm CII}/M_{gas}=0.16 \,\rm L_{\odot}/M_{\odot}$. This is consistent with our predictions for the same $G_0$ at $Z=Z_{\odot}$ (Fig. \ref{specific_cii}, upper/left panel). As $\epsilon_{\rm CII}$ is almost independent on the GMC model , in the rest of the discussion we will refer to the fiducial GMC case, i.e. model B.

\subsection{[\CII] line emission}

\begin{figure}
\centering
\includegraphics[scale=0.45]{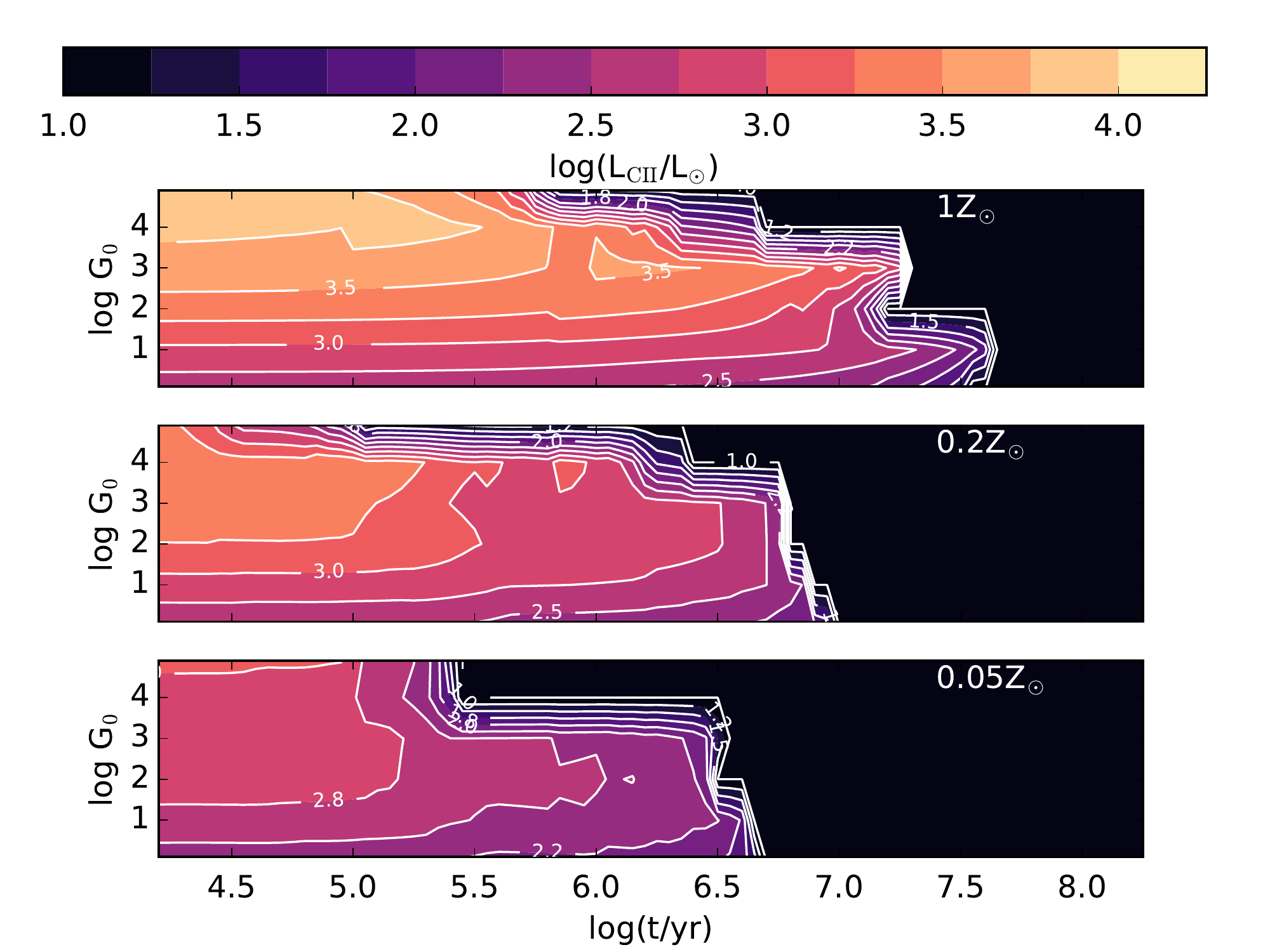}
\caption{[\CII] luminosity evolution as a function of $G_0$ for GMC model B and for different metallicities ($Z=\rm Z_{\odot}$ upper panels, $Z=0.2\,\rm{Z_{\odot}}$ middle, and $Z=0.05\,\rm{Z_{\odot}}$ lower). Black regions denote $L_{\rm CII} <10\,\rm L_{\odot}$. \label{fig_CII}}
\end{figure}

In Fig. \ref{fig_CII} we plot the [\CII] luminosity for model B as a function of $G_0$, and time, $t$, elapsed from the onset of the cloud illumination by a nearby starburst. The [\CII] line has critical densities $n^e_{crit}\approx50\,\rm{cm^{-3}}$ and $n^H_{crit}\approx 3 \times 10^3\,\rm{cm^{-3}}$ for collisions with electrons and neutral hydrogen atoms, respectively. During the PE process, the clump density always exceeds $n^H_{crit}$ (see Fig. \ref{clump_sampling2}), while the ICM density is sub-critical. Thus, ICM largely dominates [\CII] emission over the clumps. 

For $t<t_{pe}$, the [\CII] emission decreases with metallicity for strong radiation fields ($\rm log\, G_0>2$). As already mentioned, in these conditions FUV photons largely ionize GMCs; the free electrons then dominate the [\CII] excitation. However, as the ICM density exceeds the electron critical density the [\CII] luminosity is quenched.

At fixed metallicity, instead, [\CII] increases with $G_0$. The relation almost flattens for $\rm log\,G_0\geq3$ as the PDR temperature exceeds the [\CII] 158$\mu$m transition excitation temperature ($92 \,\rm{K}$; see Fig. \ref{pdr_properties}). At that point a further temperature increase does not appreciably change the [\CII] line emission \citep[e.g.][]{kaufman1999}. The flattening trend is enhanced at $Z=0.05\,\rm{Z_{\odot}}$ because $T_{\rm PDR}$ is overall higher than for $Z=Z_\odot$.
 
\begin{table*}
\centering
\caption{The coefficients of the polynomial fit for the [\CII] and [\OIII] specific luminosities at $t=10^{5}\,\rm{yr}$, as expressed in Equation \ref{polynom_cii_oiii}}.\label{cii_oiii_fit} 
\begin{tabular}{c c c c c c c}          
\hline
\hline                        
Coefficients &  & [\CII] specific luminosity & & &  [\OIII] specific luminosity &\\    
\cmidrule(l){2-4} \cmidrule(l){5-7} 
 & $Z=Z_{\odot}$ & $Z=0.2Z_{\odot}$ & $Z=0.05Z_{\odot}$ & $Z=Z_{\odot}$ & $Z=0.2Z_{\odot}$ & $Z=0.05Z_{\odot}$  \\ 
\hline                                   

    $\alpha$ & -2.424 & -2.476 & -2.652 & -2.616  & -9.079   & -9.661   \\
    $\beta$ &  0.375 & 0.505 & 0.314 &  -9.298 & 1.088  & 1.075 \\      
    $\gamma$ &  0.022 & -0.102 & -0.003  &   5.233  & 0.922  & 0.929\\
    $\delta$ & -0.009 & 0.007 &  -0.015 & -0.694 &  -0.168  & -0.168 \\
\hline                                             
\end{tabular}
\end{table*}
Note that a GMC exposed to a low $G_0$, albeit fainter in [\CII] emission, survives for a longer time. Thus, $L_{\rm CII}$ reaches a maximum at progressively later times as $G_0$ is decreased. This behavior is a characteristic imprint of the PE process, and hence is generally valid for all FIR lines (see the next Section regarding the [\OIII]). The effect of PE on the FIR line luminosity in low-metallicity GMCs can be also appreciated from Fig. \ref{fig_CII}: the $L_{\rm CII}$ after at 10 Myr drops dramatically with $Z$ as a result of PE.

A simple polynomial fit describes the specific [\CII] luminosity at $t=10^{5}\,\rm{yr}$ as a function of $G_0$:
\begin{equation}
 \label{polynom_cii_oiii}
 \begin{aligned}
{\rm log} \left(\frac{\epsilon_{\rm CII}}{L_{\odot} M_{\odot}^{-1}} \right) & = \alpha +\beta {\rm log\,}G_0 + \gamma ({\rm log\,} G_0)^2 + \delta ({\rm log\,} G_0)^3
\end{aligned}
\end{equation}
The coefficients are listed in Table \ref{cii_oiii_fit}. These results are valid up to $t_{pe}$ (see Table \ref{t_pe_table}) with a precision better than 5\% in the range $\rm log\,G_0=0-4$.

\subsection{[\OIII] line emission}
\begin{figure}
\centering
\includegraphics[scale=0.5]{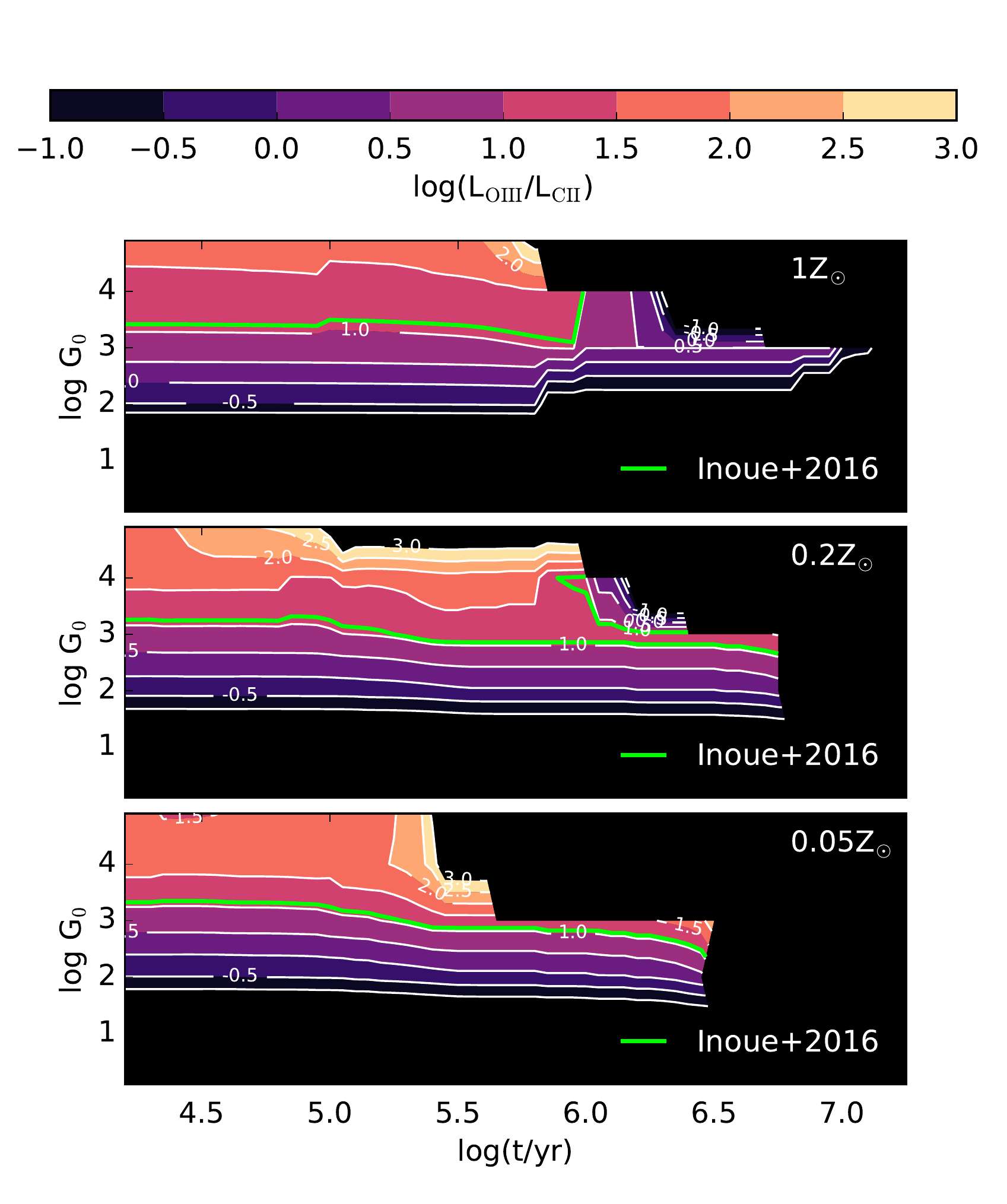}
\caption{Time evolution of the $L_{\rm OIII}/L_{\rm CII}$ ratio as a function of $G_0$ for different metallicities ($Z=Z_{\odot}$ upper panel, $Z=0.2\,Z_{\odot}$ middle, and $Z=0.05\,Z_{\odot}$ lower) for cloud model B. Black regions denote the parameter space where the ratio is either non-defined or $\log(L_{\rm OIII}/L_{\rm CII})< -1$. The green level represents the lower limit on [\OIII]/[\CII] for the Ly$\alpha$ emitting galaxy SXDF-NB1006-2 located at $z=7.2$ \citep{inoue2016}. \label{oxygen}}
\end{figure}
In this Section we extend our analysis to the [\OIII] $^3P_1 \rightarrow\, ^3P_0$ line at $88\, \rm{\mu m}$, one of the most prominent tracers of dense ionized gas at the transition with the photodissociation regions \citep[e.g][]{cormier2012}. The [\OIII] line predominantly arises from the cloud external ionized layer, as the O$^+$ ionization potential is $IP_{O^+}=35.5>13.6$ eV. Note that the critical density for collisions with free electrons is $n^e_{crit}\approx 510\,\,\rm{cm^{-3}}  \approx \langle n_{\rm ICM} \rangle$ of the model B. 

In a recent paper \citet{inoue2014} noted that ALMA capabilities might allow a combined study of the cold/neutral gas via the [\CII] line, and of the warm/ionized component using the [\OIII] $88\,  \rm \mu m$ transition well within the EoR. As a matter of fact, the [\OIII] line is often reported to be stronger than [\CII] in low-metallicity nearby dwarf galaxies \citep{cormier2012, madden2013, cormier2015}.  As a result, the [\OIII]/[\CII] may become increasingly higher at high redshift where galaxies have, on average, a lower metal content. In addition, the [\OIII] $88\,\rm{\mu m}$ line is not affected by dust extinction as it is in the case for other restframe UV/optical lines commonly used to probe \HII regions. \citet{cormier2012} showed that in local low-metallicity galaxies up to $60\%$ of the total [\OIII] 88$\rm{\mu m}$ emission originates from dense \HII regions close to the starbursts. As we only deal with GMCs, our results on the [\OIII] luminosity must be considered as solid lower bound to the luminosity of this line.

The time evolution of the [\OIII]/[\CII] ratio is plotted in Fig. \ref{oxygen}. As expected, the [\OIII]/[\CII] is correlates with $G_0$. For $\rm log\,G_0 \leq 3.5$ the ratio is almost independent of metallicity, while for $\rm log\,G_0 > 3.5$ the [\OIII]/[\CII] increases with decreasing metallicity.  If the field is weak the ionized layer is anyway thin and metallicity variations do not alter the ratio. However, for strong fields and lower metal/dust content the ionized layer becomes thick and boosts  [\OIII] line emission.  
%
However, due to the shorter photoevaporation time, at low-metallicities the maximum [\OIII]/[\CII] ratio ([\OIII]/[\CII]$=1000$ for $\rm log\,G_0>4$) can be sustained for shorter times ($t\approx 10^{6}\,\rm{yr}$ at $Z=0.2\,\rm{Z_{\odot}}$ vs. $10^{5.4}\leq t \leq10^{5.6}\,\rm{yr}$ at $Z=0.05\,\rm{Z_{\odot}}$. 

Our time-dependent calculation shows that in unevolved ($Z\leq0.2\,\rm{Z_{\odot}}$) and powerful ($\rm log\,G_0 \geq 3$) starbursts, it is possible to achieve [\OIII]/[\CII]$>10$ lasting $3-5$ Myr. This is in line with the recent findings by \citet{inoue2016} for the $z=7.2$ Ly$\alpha$ emitting galaxy SXDF-NB1006-2. In Fig. \ref{oxygen} we plot the lower limit on SXDF-NB1006-2 ([\OIII]/[\CII]$_{NB1006}>12$) for the three metallicities considered in this work. From SED fitting \citet{inoue2016} also find that $Z=0.1\,\rm Z_{\odot}$ is the most probable value for SXDF-NB1006-2, though $0.05\,\rm{Z_{\odot}}<Z<1 \,\rm{Z_{\odot}}$ cannot be rejected at a confidence level $>3\sigma$.

Finally, as in the case of the [\CII] line, in Table \ref{cii_oiii_fit} we report the coefficients of the best polynomial fit for $\rm log\,(\epsilon_{\rm OIII}/L_{\odot} M_{\odot}^{-1})$ as expressed in Eq. \ref{polynom_cii_oiii}, with a precision better than 5\% in the range $\rm log\, G_0=1-4$.
%
\section{Summary and Conclusions}\label{conclusions}
We have studied the effects of photoevaporation of GMCs irradiated by an external UV radiation field. Our model allows to compute the evolution of the GMC density field and calculate the far-infrared (FIR) line luminosity emitted by the cloud during this process, The model includes: (i) three different GMC models, with properties derived from a combination of observational and simulated results, (ii) their time evolution during the photoevaporation process.

By exploring UV field intensities (in Habing units) in the range $G_0 =1-10^{5}$ and gas metallicities $Z=1,\,0.2,\,0.05\,\rm{Z_{\odot}}$, we find that the fiducial GMC (model B in Table 1) is completely photoevaporated in a timescale $t_{pe}\le 30$ Myr.  This timescale is comparable to that deduced for destruction due to expanding \HII regions around newborn stars inside GMCs \citep{williams1997, krumholz2006}.

The PE timescale is a decreasing function of metallicity, and it goes from $30$ Myr at $Z=Z_\odot$, to 1 Myr at $Z=0.05Z_{\odot}$ for the fiducial cloud. This is because the increased penetration of FUV leads to thicker and hotter PDRs leading to a faster photoevaporation. Due to similar physical reasons, at fixed metallicity, $t_{pe}$ decreases for higher FUV fluxes. 
The presence of ionizing EUV photons becomes important for low metallicities ($Z\leq 0.2\,\rm{Z_{\odot}}$) and strong ($\rm log\,G_0 >4$) radiation fields, when the column density of the ionized layer becomes comparable to the total GMC one.

We compute the evolution of [\CII], and [\OIII] line luminosity during the PE process. {We show that the [\CII] emission per unit mass ($\epsilon_{\rm CII}$) for the three GMC models is independent of the internal GMC properties (i.e. the cloud model) and is a function of $G_0$ only, modulo a scaling factor $\propto t_{pe}$ entering the time evolution. 
It is then possible to specialize our results to the fiducial case (model B) only. 

FIR line luminosities depend on: (a) time, $t$, elapsed from onset of irradiation; (b) metallicity, $Z$, of the GMC; (c) UV field intensity, $G_0$. Albeit the interplay of these parameters is complex, a well-defined general trend emerges. Stronger UV fluxes produce higher [\CII], and [\OIII] luminosities, however lasting for progressively shorter times (i.e. $t_{pe}$ decreases along this sequence). More specifically, we find that: 

\begin{itemize}
\item[\bf 1.] For $Z=Z_{\odot}$ [\CII] emission peaks at $t \simlt 1\,\rm{Myr}$ and $\log G_0\geq 3$; the peak amplitude decreases towards lower metallicities. At fixed $Z$ the [\CII] correlates with $G_0$, even though such trend is relatively mild and tends to flatten, particularly at very low metallicity ($Z=0.05\,\rm{Z_{\odot}}$).  Note that a GMC exposed to a low $G_0$ is less luminous but its emission phase can last longer.
\item[\bf 2.] Low metallicities ($Z\leq0.2\,\rm{Z_{\odot}}$) and high UV fluxes ($\rm log\,G_0 \approx 4$) maximize the [\OIII]/[\CII] ratio, pushing it to values up to $\approx 1000$. However, due to the shorter $t_{pe}$, for very low metallicity such intense [\OIII] emission phase can be sustained only for $10^{5.4}\leq t \leq10^{5.6}\,\rm{yr}$ at $Z=0.05\,\rm{Z_{\odot}}$. 
\item[\bf 3.] The above results are consistent with recent observations of a LAE at $z\approx7.2$ \citep{inoue2016}, showing a  [\OIII]/[\CII] ratio $>12$ and $Z\approx 0.1\,\rm{Z_{\odot}}$. Under these conditions we find that gas metallicities $Z\leq 0.2\,\rm{Z_{\odot}}$ allow to sustain [\CII]/[\OIII]$=12$ for $\approx 10^{6.5}-10^{6.7}\,\rm{yr}$. 
\end{itemize} 

Although physically solid, our model has some caveats. 
As star formation within the GMC is not considered, the effects of internal radiation sources is not accounted for. Given that the star formation efficiency per free-fall time, $\epsilon_{eff}$, varies considerably in GMCs \citep[e.g.][$\epsilon_{eff}=0.1-10\%$]{semenov2015}, the estimate of the actual number of OB stars depends strongly on the local conditions of the GMC. \citet{williams1997} estimate that for clouds of mass $10^5\,\rm{M_{\odot}}$, about half are expected to contain at least one OB star.
By the way note that if an OB star forms in the GMC, the result of its ignition is to provide high UV fluxes to the cloud\footnote{The Habing flux provided by an O star at $\approx 0.1$ pc from its surface is $\log G_0\approx 6$ \citep{hollenbach1999}}. The photoevaporation timescales for a GMC destroyed by internal OB association that form blister \HII regions is $t_{pe}\approx30-40\,\rm{Myr}$ \citep{williams1997}, which is comparable to that found with our modeling at solar metallicity.\\

To conclude, we have pointed out that photoevaporation of GMCs dramatically affects their survival and FIR emission properties in a complex way. This has to be kept in mind when interpreting the FIR line data from high-$z$ galaxies, which are metal poor and characterized by hard interstellar radiation fields, all conditions leading to fast photoevaporation.  As already pointed out in \citet{vallini2015} photoevaporation feedback might be responsible for the observed spatial displacement of FIR line-emitting sites with respect to the UV continum position. In the central regions, in fact, GMC might be evaporated by the powerful radiation field, with the result that FIR lines are suppressed in the vicinity of the star-forming region. The impact of such effects on galactic scales will be explored in a forthcoming study.

\section*{Acknowledgments}
We thank the anonymous referees for their thorough and constructive comments that have greatly improved the paper. We are indebted to D. Cormier, A. Citro, E. Sobacchi, and F. Pacucci for useful comments. 
We thank all the participants of \textit{The Cold Universe} program held in 2016 at the KITP, UC Santa Barbara, for valuable comments and discussions during the workshop. This research was supported in part by the National Science Foundation under Grant No. NSF PHY11-25915. 
\bibliographystyle{mnras}
\bibliography{fdb_firlines}
\appendix
\section{Penetration of EUV and FUV photons}\label{cloudy_simulations}
When the GMC is embedded in an \HII region, the ICM is exposed to both EUV and FUV photons, while the internal clumps see only the attenuated FUV radiation. Lyman continuum photons penetrate a gas slab of density $n$, and surface area $S$, up to a depth $d$ given by:
\begin{equation}\label{ion_1}
d = \frac{\dot{N}_{ion}}{S\, n^2\, \alpha_{rec}} \approx \frac{F_{ion}}{ (h \nu)\,n^2\, \alpha_{rec}},
\end{equation}
where $\alpha_{rec}=3\times 10^{-13}\,\rm{cm^3\,s^{-1}}$ is the hydrogen recombination rate coefficient, $\dot{N}_{ion}$ the number of ionizing photons per second, $F_{ion}=L_{ion}/S$ is the ionizing photon flux, $L_{ion}$ the ionizing luminosity ($L_{ion}=\int L_{\lambda} d\lambda$ for $\lambda<912$~\AA), and $h\nu \approx 2.1\times10^{-11}\,\rm{erg}$ is the energy of Lyman limit photons. For the SED used in the present work (see Fig. \ref{sb9}), the ionizing luminosity is $L_{ion}\approx 0.3\, L_{\rm Habing}$ at 1 Myr. Hence, we can give an estimate of the column penetrated by EUV photons in term of the Habing flux as follows: 
\begin{equation}
\label{ion_2}
\begin{aligned} 
N_{\rm HII} & =n d =\frac{F_{ion}}{h\nu}\frac{1}{n\alpha_{rec}} \approx \\
& \approx 7 \times 10^{17} \frac{n}{100 \,\rm{cm^{-3}}} \frac{G_0}{\rm erg\,s^{-1}\,cm^{-2}} \,\rm{cm^{-2}}\,.
\end{aligned}
\end{equation}
For example, for $G_0=10^4-10^5$ (typical of the PDRs near OB associations) and $n=100\,\rm{cm^{-3}}$ the eq. yields $N_{\rm HII}= 7\times 10^{21} -7\times 10^{22}\,\rm{cm^{-2}}$.
In Fig. \ref{profile_ionization} we show the temperature and neutral fraction ($x_{\rm HI}=n_{\rm HI}/n$) profiles as a function of $N_H$. The profiles are obtained with \cloudy\, simulations that consider the whole SED (EUV+FUV) impinging on a gas slab with $n=100\,\rm{cm^{-3}}$ (typical of ICM) and $Z=0.05\,\rm{Z_{\odot}}$. The profiles are color-coded as a function of the Habing flux at the slab surface.

The full \cloudy \, calculation returns ionized columns $N_{\rm HII} \approx 1.5 - 7\times 10^{22}\,\rm{cm^{-2}}$ for $G_0=10^4-10^5$, i.e. in good agreement with the estimate that can be obtained from Eq.s \ref{ion_1} and \ref{ion_2} for a gas slab of pure hydrogen. 
The HII columns for a gas slab with $Z=1\,Z_{\odot}$ are $N_{\rm HII} \approx  4 \times 10^{21}\,\rm{cm^{-2}}$ for $G_0=10^4$ and $N_{\rm HII} \approx  7 \times 10^{21}\,\rm{cm^{-2}}$ for $G_0=10^5$. 
\begin{figure}
\centering
\includegraphics[scale=0.4]{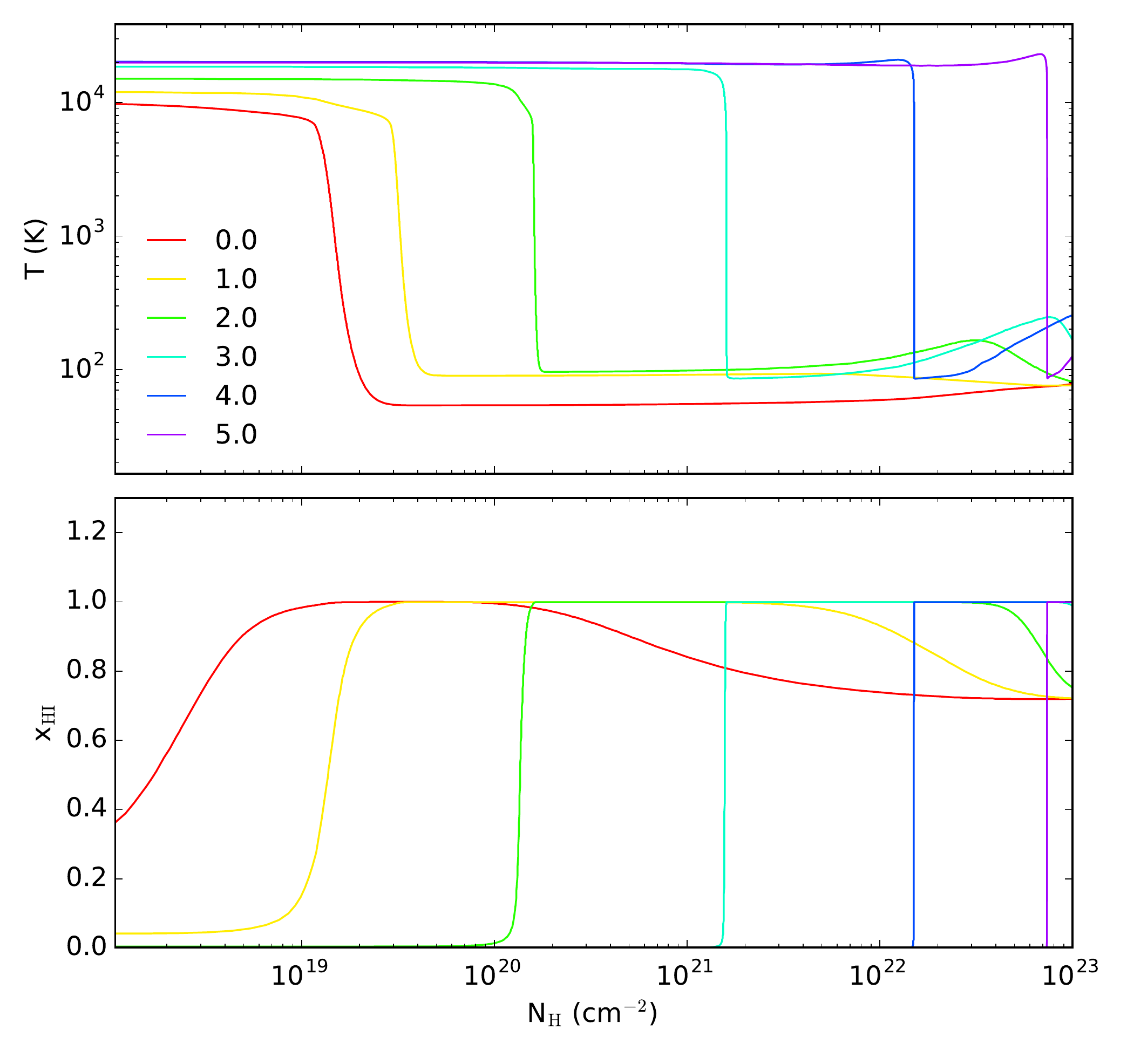}
\caption{Temperature and $x_{\rm HI}$ profiles obtained with \cloudy \,when considering a gas slab characterized by $n=100\,\rm{cm^{-3}}$ and illuminated by SED resulting from continuos star formation at 1 Myr. The lines are color coded according to $\rm log\,G_0$ at the gas slab surface. \label{profile_ionization}}
\end{figure}

Note that above such column density, any H-ionizing photons are then absorbed in a thin ($N_H\approx 10^{19}\,\rm{cm^{-2}}$ or $\Delta A_V\approx 10^{-2}$) transition zone in which the ionization structure changes from being almost fully ionized ($x_e\approx1$) to being almost fully neutral ($x_e\approx 10^{-4}$) \citep{hollenbach1999}.

In Fig. \ref{G0_profile} we plot the attenuation of the Habing flux ($G_0/G^{\rm surf}_0$) through the gas slab as a function of the metallicity. As we assume a dust-to-gas ratio that scales linearly with $Z$, at lower $Z$ corresponds a lower $G_0$ attenuation. These profiles are adopted to compute the average Habing field impinging the clump surfaces $\langle G_0^{\rm clumps} \rangle=\langle f_{att}\rangle G_0^{\rm surf}$, with:

\begin{equation}\label{fatt}
\langle f_{att} \rangle = G_{0}(\langle N^{\rm ICM-clumps}_{\rm H} \rangle)/G^{\rm surf}_0\,,
\end{equation}

where $\langle N^{\rm ICM-clumps}_{\rm H} \rangle $ is the mean column of gas in the ICM between the GMC and the clump surfaces. $\langle N^{\rm ICM-clumps}_{\rm H} \rangle= n_{\rm ICM} \langle l \rangle$ is calculated by sampling the location ($l$) of the clumps in the GMC via a montecarlo acceptance-rejection method by assuming (i) a uniform clump distribution, and (ii) the GMC to be spherical. The mean radius is $\langle l \rangle=0.25r_{\rm GMC}$.

\begin{figure}
\centering
\includegraphics[scale=0.45]{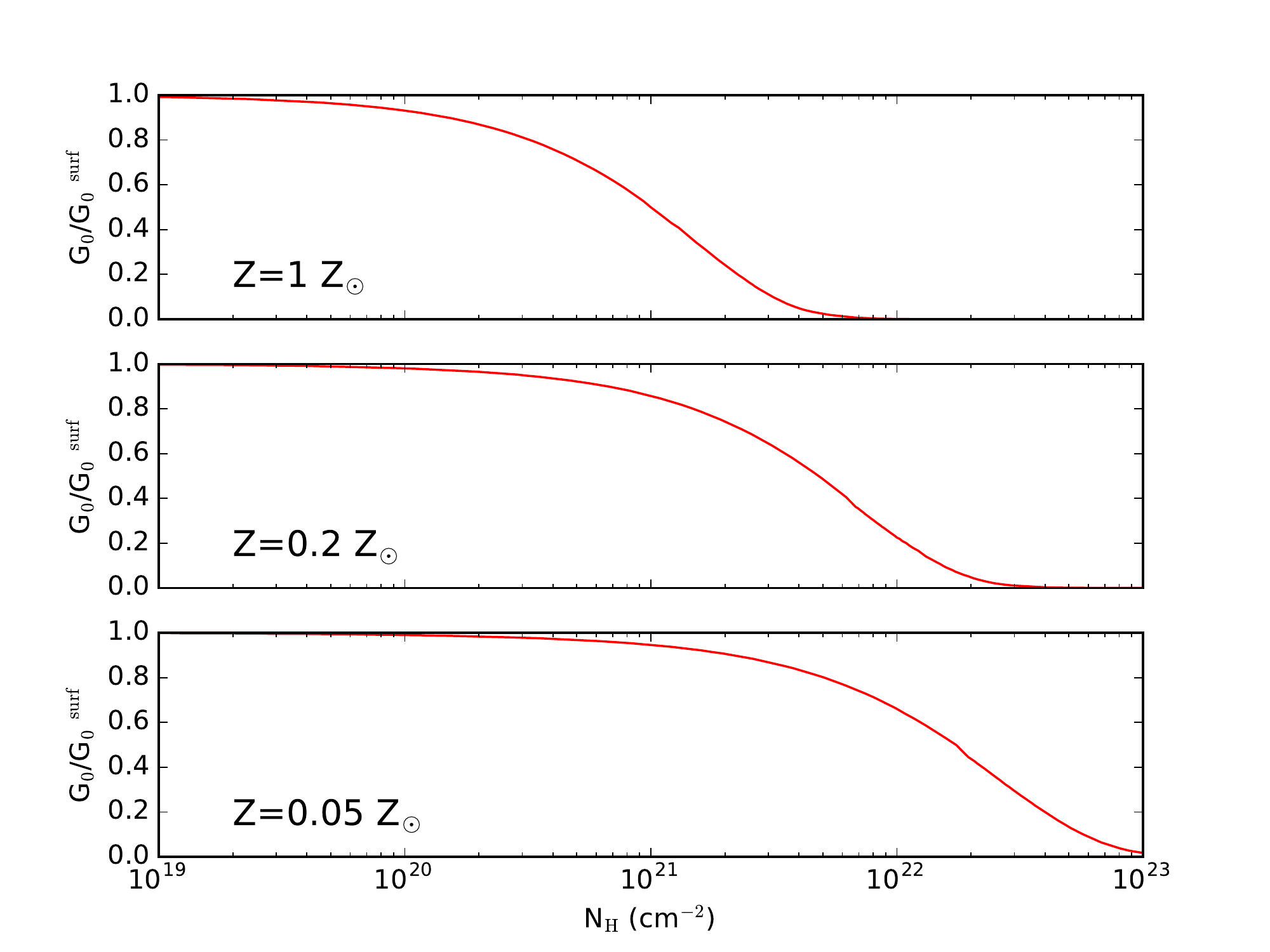}
\caption{Profile of the $G_0$ attenuation as resulting from \cloudy simulations considering a gas slab characterized by $n=100\,\rm{cm^{-3}}$ and $Z=1,\,0.2,\,0.05\,Z_{\rm \odot}$ from top to bottom. \label{G0_profile}}
\end{figure}

\section{Photoevaporation model}\label{appendices}
Clumps are assumed to be dense, small spheres of initial radius $r^0_{cl}$, and initial density $n^0_{cl}$, supported by thermal, turbulent, and magnetic pressures. The magnetic field $B$ scales with a constant power of the density so that the magnetic pressure is $P_B \propto n^{\gamma}$.
The ratios of turbulent, and magnetic pressures, over the thermal pressure are indicated with $\alpha \equiv P_{turb}/P_{T}$ and  $\beta \equiv P_{B}/P_{T}$. The fiducial values in \citet{gorti2002}, kept fixed in our work, are $\alpha = \beta = 1$ and $\gamma =4/3$.  

\subsection{Evolution of shock-compressed clumps}
Let $r_{cl}(0) = r^0_{cl} - \delta_0$, where $\delta_0 = N_0/n^0_{cl}$ is the initial thickness of PDR shell on the surface of the clump.
In the case of shock-compressed clumps, \citet{gorti2002} demonstrated (cfr. eq.s from B1 to B17 in their paper) that the shock compression shrinks the clump in a time $t_s\simeq r_{cl}(0)/c_{PDR}$ to a radius $r_s$. The clump mass at $t=t_s$ is:
\begin{equation}
m_c(t_s) = m_{cl}(0) - 8 \pi m_H N_0 r_{cl}(0)^2 \frac{v_b}{2 c_{\rm PDR}}
\end{equation}
where $v_b\approx 0.7 c_{\rm PDR}$ is the average velocity with which the radius decreases.
After being compressed by the shock, the clump radius and mass evolve as:
\begin{equation}\label{shock_r}
\begin{aligned}
r_{cl}(t>t_s) = & \left( \left[ \left[ \frac{r_s}{r_{cl}(0)} \right]^{2-1/\gamma} \right]  - \frac{2\gamma -1}{3\gamma -1} \frac{6\nu \eta_{0}}{(\eta_0-1)^2} \times \right. \\
& \left. \left[ \frac{\beta (\eta_0 -1)}{2 (2\nu^2 + \alpha)} \right]^{1/\gamma} \left[ \frac{t}{t_c} - \frac{(\eta_0 -1)}{\eta_0 \nu} \right] \right)^{\gamma / (2 \gamma - 1)}
\end{aligned}
\end{equation}
and
\begin{equation}\label{shock_m}
\begin{aligned}
m_{cl}(t>t_s) = m_{cl}(0) \left[ \frac{2 (2\nu^2 + \alpha)}{\beta (\eta_0 -1)^2} \right]^{1/\gamma} \left[ \frac{r_{cl}(t)}{r_{cl}(0)} \right]^{3-1/\gamma}.
\end{aligned}
\end{equation}
The photoevaporation timescale, obtained by setting the clump radius to zero, is
\begin{equation}\label{shock_tpe}
\begin{aligned}
t_{pe}= & t_c \left[\left [ \frac{r_s}{r_{cl}(0))} \right ] ^{2-1/\gamma} \left( \frac{3\gamma-1}{2\gamma-1} \right) \times \right. \\
 & \left. \times \frac{(\eta_{0} -1)^2}{6\nu\eta_{0}} \left [ \frac{2(2\nu^2+\alpha)}{\beta(\eta_{0} -1)} \right ]^{1/\gamma} + \frac{\eta_{0}-1}{\eta_{0}\nu} \right] \\
\end{aligned}
\end{equation}

\subsection{Evolution of clumps with initial expansion}
Assuming that the clumps expands in the vacuum \citep[cfr. eq.s from C1 to C7 in ][]{gorti2002} at their sound speed $c_c$ until the pressure drops to that in the heated outer layer, it is possible to demonstrate that the expansion time is:
\begin{equation}
t_e = t_c \left(1-\frac{1}{\eta_0}\right) \left[ \left[ \frac{3\nu + \eta_0 - 1}{3\nu 2(2\nu^2 + \alpha)/(1+\alpha)} \right]^{1/2} -1 \right].
\end{equation}
The time evolution of the clump radius and mass at $t>t_e$ is:
\begin{equation}
r_{cl}(t>t_e) = r_{cl}(t_e) - \frac{3}{2} c_{\rm PDR} \frac{1+\alpha}{2\nu^2 + \alpha} (t-t_e),
\end{equation}
and
\begin{equation}
m_{cl}(t>t_e) = m^0_{cl} \frac{2(2\nu^2 +\alpha)}{\eta_{0} (1+\alpha)} \left[ \frac{r_{cl}(t)}{r^0_{c}} \right]^2.
\end{equation}
The photoevaporation timescale is obtained by setting the clump radius to zero
\begin{equation}\label{exp_tpe}
\begin{aligned}
t_{pe}= \left( \frac{2 \nu^2 +\alpha}{1+\alpha} \right) \frac{r_{cl}(0)+ c_c t_e}{3c_{PDR}} + t_e 
\end{aligned}
\end{equation}
\bsp	
\label{lastpage}
\end{document}